\documentclass[final,5p,times,twocolumn]{elsarticle}

\usepackage{amssymb}
\usepackage{amsmath}

\usepackage{amsmath,amssymb,amsfonts}
\usepackage{algorithmic}
\usepackage{graphicx}
\usepackage{textcomp}
\usepackage{xcolor}
\usepackage{hhline}
\usepackage{pifont}
\usepackage{svg}
\usepackage{multirow}
\usepackage{datetime}
\usepackage{tabularx}
\usepackage{threeparttable}
\usepackage{booktabs}
\usepackage{makecell}
\usepackage{float}
\usepackage{subcaption}
\usepackage{hyperref}
\usepackage{xurl}
\usepackage{tablefootnote}

\usepackage{natbib}
\biboptions{sort&compress}

\newcommand{\ra}[1]{\renewcommand{\arraystretch}{#1}}

\journal{}

\begin{document}

\begin{frontmatter}



\title{An Overview and  Solution for Democratizing AI Workflows at the Network Edge}



\author{Andrej \v{C}op\corref{cor1}}
\cortext[cor1]{Corresponding author.}
\ead{andrej7.cop@gmail.com}

\author{Bla\v{z} Bertalani\v{c} and Carolina Fortuna}

\address{Jozef Stefan Institute, Ljubljana, SI-1000, Slovenia}

\begin{abstract}
With the process of democratization of the network edge, hardware and software for networks are becoming available to the public, overcoming the confines of traditional cloud providers and network operators.
This trend, coupled with the increasing importance of AI in 6G and beyond cellular networks, presents opportunities for innovative AI applications and systems at the network edge.
While AI models and services are well-managed in cloud systems, achieving similar maturity for serving network needs remains an open challenge.
Existing open solutions are emerging and are yet to consider democratization requirements.
In this work, we identify key requirements for democratization and propose NAOMI, a solution for democratizing AI/ML workflows at the network edge designed based on those requirements.
Guided  by the functionality and overlap analysis of the O-RAN AI/ML workflow architecture and MLOps systems, coupled with the survey of open-source AI/ML tools, we develop a modular, scalable, and distributed hardware architecture-independent solution.
NAOMI leverages state-of-the-art open-source tools and can be deployed on  distributed clusters of heterogeneous devices.
The results show that NAOMI performs up to 40\% better in deployment time and up to 73\% faster in AI/ML workflow execution for larger datasets compared to AI/ML Framework, a representative open network access solution, while performing inference and utilizing resources on par with its counterpart.

\end{abstract}




\end{frontmatter}



\section{Introduction}

Democratization of the network edge aims towards a widely accessible edge cloud for computation and communication while not restricting strictly to the domain of cloud providers and network operators. 
With open hardware, e.g. USRP, BladeRF, LimeSDR, and software, e.g. Open5GS, OAI-5GCore, srsRAN, for 5G  networks becoming available to the public~\cite{amini20235g, mihai2022open}, democratization is lowering the boundary for anyone to deploy and operate cellular networks, including the access network and edge clouds.
Further, companies outside of telecommunication are deploying private 5G networks for automation use cases with a demand for low latency connectivity to intelligence nearby providing an opportunity for the development of democratized AI applications and systems for the network edge~\cite{Peterson2019DemocratizingEdge}. Such an open ecosystem has the potential to support increased innovations and business models significantly boosting efficiency ~\cite{aijaz2020private}.

AI is also foreseen to be a centerpiece in 6G and beyond cellular networks~\cite{Letaief2019TheNetworks} with some traditionally networking functionality being replaced by AI based realization towards so-called AI native functionality~\cite{hoydis2021toward,wu2022ai}. AI algorithms are envisioned to achieve intelligent, sustainable, and dynamically programmable services supporting the adaptability and flexibility of the network~\cite{Letaief2019TheNetworks}. For instance, the PREDICT-6G~\cite{predict6g} project deploys AI models to the control plane where predictability algorithms are required to proactively allocate 6G network resources.

O-RAN (Open Radio Access Network) Alliance is tackling the standardization of the access segment of such open networks by defining the architectural design and specifications. As part of O-RAN, also the AI/ML workflow is defined, including components, scenarios, and requirements for such systems to operate on radio access networks~\cite{b1}.
Furthermore, the O-RAN SC (Software Community), which is a collaboration between O-RAN Alliance and Linux Foundation, builds open-source software as defined in the specifications by O-RAN Alliance and provides a solution for AI/ML workflow system - AI/ML Framework~\cite{o-ran-sc-docs}. On the other hand, automated AI/ML workflows, also referred to as 
MLOps (Machine Learning Operations) ~\cite{Diaz-De-Arcaya2023ASurvey} in the recent years, are well established in cloud infrastructures~\cite{li2014scaling}. Additionally, large and mature open source technologies and tools, powering production systems, have been developed for the various needs of such pipelines~\cite{zaharia2018accelerating,mlflow,bentoml,moritz2018ray,optuna_2019}.

While AI models and services are already scaled and well managed in cloud systems as recently shown with the plethora of large language models, reaching the same level of maturity for serving the needs of networks is still an open research and engineering endeavor~\cite{liu2024operationalizing}.
The existing software solves some challenges however lacks in maturity and ease of use along with the implementation of design mechanisms of democratization such as heterogeneity, customization, scalability, and distributed services~\cite{Peterson2019DemocratizingEdge}. For instance, the current O-RAN solution is also partly powered by cloud native open source components: for data preparation the Influx~\cite{influxdatawebsite} time series database~\cite{esling2012time} and the Cassandra~\cite{lakshman2010cassandra}  distributed data management system were selected, Kubeflow~\cite{kubeflow}, a ML model development and training system~\cite{zaharia2018accelerating} was selected for model training and monitoring while Kserve~\cite{kserve} was selected for model serving. However, \textit{it is unclear what were the design and performance decisions leading to the existing O-RAN solution technology and tool  selection} as well as the need for the development of  custom components such as the feature store and training manager.

Furthermore,  O-RAN and Mobile Edge Computing (MEC) are both significant but complementary concepts in the evolving landscape of cellular communications and there are pathways for their integration~\cite{chih2020perspective}. Therefore, \textit{enabling seamless instantiation and management of AI/ML workflows at the network edge} and subsequently positioning specialized models closer to actors, while allowing the deployments to scale and offload on cloud devices could create new economies. This way, stakeholders from the telecommunications but also outside could develop new services for digital transformation and increased infrastructure  automation~\cite{Peterson2019DemocratizingEdge}.

In this paper, we start by understanding general requirements for democratizing software and services with a focus on AI/ML lifecycle management. Then we investigate existing approaches, technologies and open-source tools for managing AI/ML workflows in the access network~\cite{o-ran-sc-docs} and the cloud~\cite{diaz2023joint} and propose a new solution that enables their democratization and better model management through self-evolving principles~\cite{Dellarocas1998ArchitectureSystems}. The contributions of this paper are as follows.

\begin{itemize}
    \item  The identification of requirements for democratized AI/ML workflows at the edge based on existing software and systems works as well as the need for automated workflow management. 
    \item Functionality and overlap analysis of the AI/ML workflow in the O-RAN specification and MLOps systems from the cloud computing community, coupled with the survey of open-source tools for AI/ML workflows. The analysis serves as a guideline to understand the current state of AI/ML workflow automation at the network edge and supports the selection of the appropriate tools according to specific requirements therefore speeding up the process of creating an AI/ML solution.
    \item Propose \textit{NAOMI}\footnote{\url{https://github.com/copandrej/NAOMI}}, a solution for democratizing AI/ML workflows at the network edge. NAOMI is modular, scalable, distributed, architecture-independent and easy to use while leveraging SotA open-source tools. We show that the system can be deployed on  distributed clusters of heterogeneous devices.
    \item The results show that NAOMI performs up to 40\% better in deployment time and up to 73\% faster in AI/ML workflow execution for larger datasets compared to AI/ML Framework, a representative open network access solution, while performing inference and utilizing resources on par with its counterpart.
\end{itemize}

The remainder of this paper is organized as follows. 
In Section~\ref{sec:related_work} we discuss the related work, while in Section~\ref{sec:requirements} we identify the requirements of democratized AI/ML workflow systems.
In Section~\ref{sec:workflow_architecture} we present the architecture of AI/ML workflows and in Section~\ref{sec:tools} compare open-source tools for AI/ML components. 
According to this analysis, we propose a solution for the democratized AI/ML system in Section~\ref{sec:solution}, explain the methodology for evaluation in Section~\ref{sec:methodology}, and present evaluation results when comparing it to the existing O-RAN solution in Section~\ref{sec:results}. 
Finally, we conclude the paper in Section~\ref{sec:conclusion}.

\section{Related Work} \label{sec:related_work}
Democratizing technology refers to cultural practices in which more people are able to engage with it in an easier way~\cite{tanenbaum2013democratizing}. Also in software and system engineering, a number of works analyzed existing platforms able to support democratization and proposed new ones. For instance ~\cite{Peterson2019DemocratizingEdge} defined the democratization of the network edge, outlined the opportunities and mechanism design for democratizing the network edge, and proposed an experimental platform for access-edge cloud. Furthermore,~\cite{rall2023towards} acknowledged that "recent AI research has significantly reduced the barriers to apply AI, but the process of setting up the necessary tools and frameworks can still be a challenge". They identified the key requirements for a platform that can achieve true democratization of AI and highlight the need for self-hosting options, high scalability, and openness. In this work, inspired by some of the requirements in~\cite{Peterson2019DemocratizingEdge,rall2023towards}, the recent push to open up the radio access part of the networks enabled by O-RAN ~\cite{o-ran-sc-docs,Polese2023UnderstandingChallenges}, and the streamlining of data cleaning and pre-processing, model training and evaluation through MLOps~\cite{diaz2023joint,Kreuzberger2023MachineArchitecture}, we further the knowledge on the democratization of AI for the network edge and develop a new, open, easy to use and modular framework based on user-friendly, stable open source tools and evaluate it against the existing O-RAN AI/ML Framework solution.

The O-RAN Software Community (O-RAN SC) is developing an open-source solution for AI/ML workflows for O-RAN also referred to as AI/ML Framework~\cite{o-ran-sc-docs}. 
Lee et al. utilized this framework, analyzed shortcomings with model packaging and scalability, and suggested improvements in~\cite{Lee2020HostingPlatform}.
Further, they provided a solution for personalized network optimization via reinforcement learning in~\cite{Lee2021O-RANLearning}, utilized O-RAN SC software, and showcased the benefits of intelligent control of RAN. While performing a detailed analysis of the O-RAN architecture and interfaces, Polese et al.~\cite{Polese2023UnderstandingChallenges} also briefly reflected on the AI/ML workflow architecture and a use case example.
Most other related works, such as Giannopoulos et al.~\cite{Giannopoulos2022SupportingCases},  Sun et al.~\cite{Sun2024IntelligentBeyond}, Marinova et al.~\cite{Marinova2024IntelligentOpportunities}, and Bonati et al.~\cite{Bonati2023OpenRANPlatforms} focus on the use cases, sometimes including intelligent models provided by the workflow. 
Similarly to AI/ML workflows, RAN slicing is an important part of modern networks, allowing the creation of virtual networks tailored to different needs~\cite{o-ran-slicing}. 
Foukas et al.~\cite{Foukas2017Orion:Architecture} introduced a framework ORION for RAN slicing that provides isolation between slices while using resources efficiently and Chen et al.~\cite{Chen2023Channel-AwareSchedulers} focused on improving scheduling by making it channel-aware at both inter-slice and enterprise levels. 
While we do not contribute directly to RAN slicing, we recognize its role, along with AI/ML, in building flexible network systems.
In contrast, our work performs an in-depth analysts of the O-RAN AI/ML workflow from the O-RAN Alliance Work Group 2~\cite{b1} also in relation to MLOps and then further focuses on proposing a new, democratized system for AI/ML workflows for the network edge.

While the MLOps term was coined in the cloud computing community~\cite{diaz2023joint,Kreuzberger2023MachineArchitecture}, a number of papers employ the related terminology also for networks and IoT, for instance, Rezazadeh et al.~\cite{10475843} proposed SliceOps, an MLOps system focused on automation-native 6G networks. 
They utilized explainable AI and MLOps to automate AI lifecycle for network slicing. Raj et al.~\cite{Raj2021EdgeApplications} proposed an automation framework for AIoT applications called Edge MLOps.
They utilized cloud resources for the orchestration of ML training and data storage and edge devices for ML inference.
They implemented automated ML to evaluate model drift and retrain the ML models. While Rezazadeh et al.~\cite{10475843} focused on 6G network slicing and Raj et al.~\cite{Raj2021EdgeApplications} relied on the Azure proprietary platform, our work is edge application agnostic focusing on openness, simplicity, modularity and other requirements specific to democratized technology.
 
There are multiple papers related to MLOps that focus on tool surveys and architecture definitions.
Ruf et al.~\cite{Ruf2021DemystifyingTools} provided a detailed recipe for the selection of open-source tools for MLOps.
Kreuzberger et al.~\cite{Kreuzberger2023MachineArchitecture} performed a literature and tool review as well as an interview study regarding MLOps, while Ashmore et al.~\cite{Ashmore2021AssuringChallenges} performed a survey of MLOps architecture, however using different definitions to other related papers.
Other papers~\cite{Symeonidis2022MLOpsChallenges, Cerar2023FeatureNetworks, Recupito2022AFeatures} provide definitions and tools for MLOps.
We expand upon the MLOps-related papers by providing a comparative analysis with O-RAN and a related open-source tool comparison completed by the proposed democratized framework evaluated  against the most recent O-RAN solution on on-premise general-purpose and edge devices.

\begin{figure*}[htbp]
    \centering
    \begin{subfigure}[b]{0.55\linewidth}
        \centering
        \includegraphics[width=\linewidth]{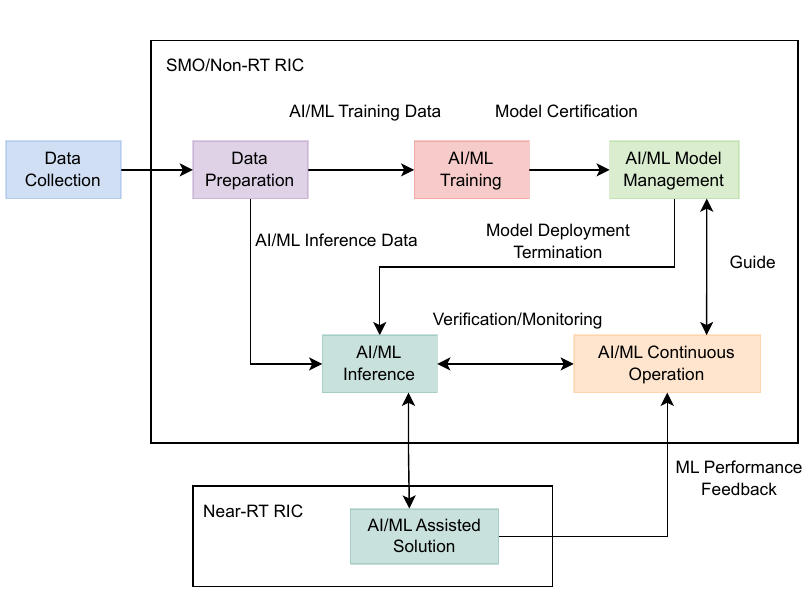}
        \caption{AI/ML workflow on O-RAN architecture, adapted from~\cite{b1}.}
        \label{subfig:oran_architecture}
    \end{subfigure}
    \par\bigskip  
    \begin{subfigure}[b]{0.55\linewidth}
        \centering
        \includegraphics[width=\linewidth]{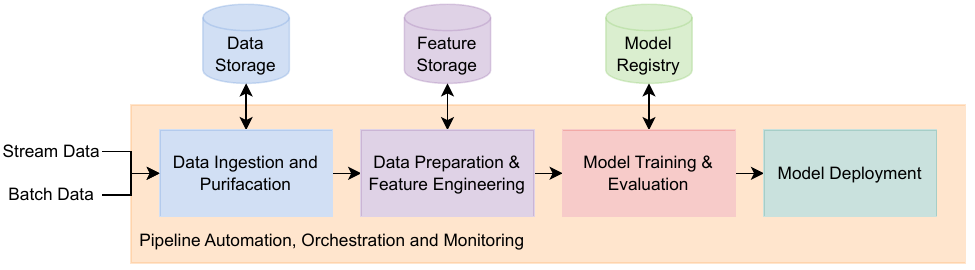}
        \caption{A general MLOps pipeline~\cite{Cerar2023FeatureNetworks, Ruf2021DemystifyingTools, Kreuzberger2023MachineArchitecture, Google2020MLOps:Learning}.}
        \label{subfig:mlops_pipeline}
    \end{subfigure}
\end{figure*}

\section{Requirements for AI/ML Workflow Democratization} \label{sec:requirements}

The democratization of AI/ML workflows represents a shift towards accessibility and efficiency in the deployment and utilization of AI/ML technologies. 
In general, democratization involves making advanced technologies accessible and usable by a broad audience, not just experts or those with specialized skills~\cite{tanenbaum2013democratizing,Peterson2019DemocratizingEdge}. 
Central to this is the establishment of a system that democratizes access to AI/ML capabilities~\cite{rall2023towards} and ensures seamless deployment across diverse network and on-premise infrastructure. Starting with the fundamental principles of democratized network edge  presented in~\cite{Peterson2019DemocratizingEdge} and the requirements towards democratizing AI from~\cite{rall2023towards}, we identify the following as relevant requirements for AI/ML workflows at the edge.

\subsection*{Openness}
Openness is crucial in a democratized system because it enables transparency, encourages collaboration, and allows for community-driven innovation. In the case of software systems, openness translated in the utilization of open-source software with sufficiently permissive licenses, sufficiently large developer community and activity. This ensures cost-effective operation and maintenance of the system and promotes open access networks. 

\subsection*{Ease of use}
For any democratized solution, the emphasis lies in enabling straightforward deployment and accessibility. This implies sufficiently clear documentation, set-up via a few simple commands or automated through scripts and support of several operating system and tool versions.  
Users with basic infrastructure and fundamental computer engineering knowledge should be able to deploy, customize, and utilize the system effectively. 

\subsection*{Modularity}
Certain users of the system may not require all available services or may prefer to switch components to suit their specific use cases. 
Therefore, the solution must be modular, allowing components to be interchanged with minimal engineering effort, possible through a simple configuration file adaptation. 
Each component should function independently, enabling users to utilize each part of the system individually.


\subsection*{Self-Evolving}
The self-evolving requirement is not considered in conventional work related to democratized systems~\cite{rall2023towards,Peterson2019DemocratizingEdge}, however we deem it important in the context of AI where the system should also offer support for self-evolving models, i.e. automatic model management and retraining.  
Generally, self-evolving software can adapt dynamically to changing circumstances or internal conditions.
The AI/ML workflow system should support metrics collection and automated, scheduled triggers for workflows to retrain and redeploy ML models as self-evolving AI/ML services~\cite{Dellarocas1998ArchitectureSystems} upon conditions that may change the expected performance of the deployed model.

\subsection*{Heterogeneity and Virtualization}
Designs of democratized solutions for network edges must accommodate heterogeneity, as edge devices vary in CPU architecture and hardware. 
Given that modern networks are software-defined and virtualized, with network functions managed by orchestration platforms~\cite{Li2017IntelligentIntelligence}, a solution that enables democratization needs to be containerized, orchestrated, and decoupled from specific platforms, while also being amenable to more resource constrained devices typical of some edge and far edge environments.

\subsection*{Distributed Services and Scalability}
Support for distributed services is crucial in multi-node network edges, however distributed training and inference is a supplementary capability to balance the computational load on a multiple nodes and reduce the latency. 
Effective orchestration of tasks across a distributed cluster, possibly formed of heterogeneous devices, is essential for good performance and quality of service.

\section{Workflow Analysis: O-RAN AI/ML vs MLOps} \label{sec:workflow_architecture}

Various definitions, requirements, and components exist for MLOps (Machine Learning Operations) pipelines and AI/ML workflows. In this section, we analyzed AI/ML workflow components as specified in the latest AI/ML specifications of O-RAN Alliance's Work Group 2 released in July 2021~\cite{b1} and depicted in Fig.~\ref{subfig:oran_architecture}, and compare them to general MLOps pipelines and its corresponding phases presented in Fig.~\ref{subfig:mlops_pipeline}~\cite{Cerar2023FeatureNetworks, Ruf2021DemystifyingTools, Kreuzberger2023MachineArchitecture, Google2020MLOps:Learning}. 

According to the AI/ML workflow specifications of the O-RAN Alliance, a general framework includes the following ML components: data preparation, model training, model management, model inference, and continuous operation~\cite{b1}.
These components collectively form a workflow (pipeline) with a set of AI/ML functionalities with input and output artifacts for each component as shown in Fig.~\ref{subfig:oran_architecture}. While less convoluted in terms of the arrangement of the functional blocks and their interconnection, it can be seen from the MLOps pipeline from Fig.~\ref{subfig:mlops_pipeline}~\cite{Cerar2023FeatureNetworks, Ruf2021DemystifyingTools, Kreuzberger2023MachineArchitecture, Google2020MLOps:Learning} that it contains similar components as the top one. For instance, coloured with blue at the top the data collection functional block can be observed, while colours with blue at the bottom, the stream and batch data ingestion, subsequent purification and storage components are depicted. The purple-coloured data preparation functional block at the top corresponds to the data preparation, feature engineering and feature storage blocks at the bottom. Similar color mapping can be observed for the remaining components while as more detailed explanation of the necessity of the respective functional blocks and their similarity is provided in the following paragraphs. 


Depending on the use case, certain AI/ML workflow components can be placed in O-RAN's Non-RT RIC (Non Real Time RAN Intelligent Controller) or Near-RT RIC (Near Real Time RAN Intelligent Controller) with the RIC being an element of the SMO (Service Management and Orchestration) framework.
RIC is responsible for controlling RAN functions and the Near-RT RIC provides a lower latency for actions as it resides on edge or regional cloud~\cite{Polese2023UnderstandingChallenges}.
The AI/ML assisted solution is a deployed application (usually on Near-RT RIC as presented in the mentioned Fig.~\ref{subfig:oran_architecture}), which is an AI solution for RAN's functionalities and with RIC provides control and optimization of RAN elements and resources.
Data from RAN is collected and used for offline/online learning and inference in the AI/ML workflow~\cite{b1}. An example use case would be controlling RAN slicing by adapting the ML model according to the current network load as considered in~\cite{Polese2023UnderstandingChallenges}.
The workflow consists of collecting data and metrics over the RAN interfaces and generating datasets.
The model is selected and trained on the dataset and deployed on the Near-RT RIC.
Slicing control is performed using inference output on KPMs (Key Performance Metrics) and if any anomalies or inefficiencies are detected with the model, retraining can be triggered.

\begin{figure*}[htbp]
\centerline{\includegraphics[width=0.6\linewidth]{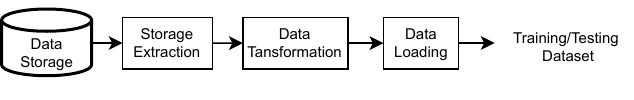}}
\caption{AI/ML data preparation ETL process.}
\label{fig:data-prep}
\end{figure*}

A solution for AI/ML workflow was released by the O-RAN Alliance Software Community, which collaboratively develops open-source software for the RAN in conjunction with the O-RAN Alliance and Linux Foundation.
O-RAN SC creates software aligned with the O-RAN architecture and specifications.
The AI/ML Framework is one of the O-RAN SC projects that supports features in accordance with the O-RAN Alliance WG2 specification.
The AI/ML Framework project is still under development and in its current version, it consists of multiple modules for data extraction, model training, and feature store and utilizes open-source software such as Kubeflow, KServe, InfluxDB, Leofs and Cassandra to realize AI/ML workflow components~\cite{o-ran-sc-docs, Lee2020HostingPlatform}.
At the time of writing this paper version \textit{1.2.0 I Release} is officially released.
As the software is in early development, some of the features and functionalities are yet to be documented, thus for this work we considered sources from official documentation~\cite{o-ran-sc-docs}, research papers~\cite{Lee2020HostingPlatform, Lee2021O-RANLearning}, and official Confluence wiki~\cite{ai-ml-framework}.

\subsection{Data Preparation Component}
\label{subsec:components:data_prep}

In the latest release of the O-RAN Alliance specification, the data preparation component is defined as a pipeline, which is referred to as an ETL process (extract-transform-load).
This process is presented in Fig.~\ref{fig:data-prep} and is part of the AI/ML workflow architecture depicted in Fig.~\ref{subfig:oran_architecture}.
It consists of storage and storage extraction, data transformation, and data loading, where the data is split into a training dataset and a testing dataset.
Data analysis, where data quality assessment is performed, data validation, and data cleaning (i.e. purification) stages, also depicted in the MLOps pipeline in Fig.~\ref{subfig:mlops_pipeline}, can be part of the data preparation pipeline.
The storage location for the data can be a storage bucket, a generic storage volume or a data lake~\cite{b1}.
Within the AI/ML Framework project, a solution for data extraction as well as a Feature Store SDK was developed.
In the latest release of their solution, they utilize InfluxDB as a data lake.
They create features and store them with a Feature Store SDK interface in Apache Cassandra~\cite{aiml-fw-athp-data-extraction}.

In general MLOps requirements data ingestion, data analysis, and data transformation are the main functions of the data preparation component~\cite{Recupito2022AFeatures}.
The data storage and data acquisition process, often referred to as \textit{Big Data}, is an initial collection of data and storage along with data processing and depending on the definition can be a part of MLOps or a separate data preparation process~\cite{bigdata, Ruf2021DemystifyingTools}.
In some cases, data versioning, typical for data science and data management, can also be used to track changes made to the data over time.
Versioning is commonly used in development environments as it ensures data reproducibility and enables collaboration~\cite{mlops-guide-data}, however in production environments it may bring unnecessary overhead.

Data storage is important in a production environment to fully automate the collection and storage of data.
After the data is collected it needs to be stored in a persistent manner either structured and processed or stored for later processing.
As a simplest solution with the lowest overhead, object storage solutions such as storage buckets are used.
For structured querying and processing,  data lakes and data warehouses are utilized and for the highest performance structured databases can be used.

Feature engineering and feature management are part of data preparation in MLOps pipelines, as illustrated with purple color in Fig.~\ref{subfig:mlops_pipeline}. 
Depending on the terminology feature engineering can have the same function as data transformation.
A feature store tool contains instructions on how features are defined and how to produce them.
It ingests data, transforms it, and stores it in a feature store registry.
Features are then available for the model training stage and prediction~\cite{Cerar2023FeatureNetworks}.

The last step is data loading, which consists of splitting data into training and testing datasets for model training.

\subsection{AI/ML Training Component}
\label{subsec:components:training}

\begin{figure*}[htbp]
\centerline{\includegraphics[width=0.7\linewidth]{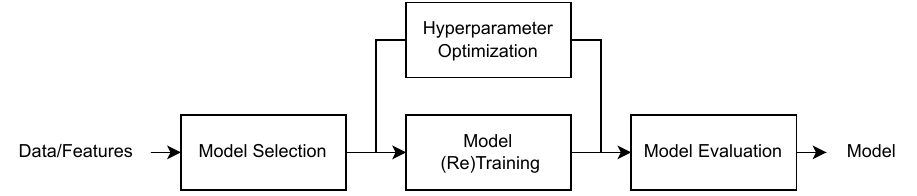}}
\caption{AI/ML training pipeline, adapted from~\cite{b1}.}
\label{fig:train}
\end{figure*}

O-RAN's model training procedure consists of the selection and training of an ML model in relation to an AI/ML assisted solution.
Once the model is trained, evaluated, and validated, it is published to the model catalog (AI/ML model management).
The model training phases include model selection, model training, hyperparameter optimization, and model evaluation.
These phases form a training pipeline with input data from the data preparation component and produce an ML model, as shown in Fig.~\ref{fig:train}.
AI/ML training component fits between data preparation component and model management in O-RAN AI/ML workflow architecture, as depicted in Fig.~\ref{subfig:oran_architecture}.
Furthermore, this component supports model retraining triggered by continuous operation and performance feedback.
Existing models can then be fetched from the model catalog for retraining on updated data and the new version can be updated to the catalog for the model inference component.
In some O-RAN use cases, online training on Near-RT RIC is performed, which is a process of improving an AI/ML model where the model learns incrementally as new data arrives sequentially.
According to the O-RAN Alliance, this type of training should only be used to fine-tune and update a model that was trained offline first~\cite{b1, Polese2023UnderstandingChallenges}.

In the latest release of O-RAN SC AI/ML Framework, the AI/ML training component is realized with a training manager on an AI training host platform that trains on data extracted with the Feature Store SDK from the Casandra database.
The model training can be controlled via the dashboard portal and the progress of the training can be monitored.
Training jobs can be created on the dashboard by selecting the data source, feature group, and training function.
Training jobs are versioned and the status of training jobs is reported on the Dashboard.
The solution only supports the management and hosting of AI/ML training.
Machine learning libraries have to be utilized for training and evaluation of models~\cite{ai-ml-framework}.

MLOps definitions of model training are similar to O-RAN's, with model training and evaluation situated after the data preparation component in Fig.~\ref{subfig:oran_architecture} in O-RAN and Fig.~\ref{subfig:mlops_pipeline} in MLOps.
Model training is the core part of any ML project.
In MLOps, first, an appropriate model and model design need to be selected.
Data from the feature store or from the data preparation component is used for training and evaluating the model.
CT (Continuous Training) is a practice in MLOps for retraining the model when the performance of the model degrades.
In the case of edge infrastructure or distributed systems, a framework that trains models on multiple machines is needed.
Automated hyperparameter optimization is an important part of the ML model training, which can be done by a specialized tool for hyperparameter optimization or by a model training library.
Optionally, training metrics are collected, and experiment information is visually reported by an experiment tracker~\cite{Ruf2021DemystifyingTools}.

Model selection is a process of selecting the appropriate ML model depending on data and the required solution.
Multiple models can be combined to create a model ensemble to improve performance.
Some automated tools exist for model selection, for tabular data, AutoML or Neural Architecture Search (NAS) tools can choose the appropriate model, but in general, model design and selection are manual processes~\cite{He2021AutoML:State-of-the-art, Elsken2019NeuralSurvey}.

Hyperparameter tuning is an automated process of selecting the best model training parameters to produce an accurate model.
It is a computationally demanding task involving iterating through parameters, retraining, and reevaluating the model~\cite{yu2020hyperparameteroptimizationreviewalgorithms}.

ML model training involves learning values for model weights to create a prediction function and increase the accuracy of the model.
The implementation of training differs greatly depending on the type of learning and algorithm, and the specifics are outside the scope of MLOps and this paper.
Common ML libraries like TensorFlow, PyTorch, and Scikit-learn are used to train the model~\cite{Gevorkyan2019ReviewLearning}.
Model training duration can be reduced by using specialized processing units like GPUs (Graphics Processing Units) and TPUs (Tensor Processing Units) or by distributing the computation across multiple nodes~\cite{Steinkraus2005UsingAlgorithms, Jouppi2017In-DatacenterUnit}. 

Distributed training is applicable for multiple machines working in a cluster.
There are two approaches to distributed training.
A data-parallel approach where data is divided between nodes or a model-parallel approach where parts of the model are trained on multiple machines and results can be aggregated to produce a complete model faster~\cite{Verbraeken2020ALearning}. During training and after the model is produced, the model is evaluated on the testing dataset to assess performance.

Experiment tracking and monitoring are features especially useful in a development environment to track the progress and metrics of model training.
Metrics can then be visually reported on a dashboard and graphs.
Some training metrics reporting is expected in fully automated workflows to monitor the continuous operation of training and ensure optimal model performance~\cite{Ruf2021DemystifyingTools}.

\subsection{AI/ML Model Management}
\label{subsec:components:model_management}
The O-RAN Alliance defines model management as the network function responsible for handling the deployment of machine learning models on the inference host.
The model management component receives models from the AI/ML training component as depicted in AI/ML workflow architecture in Fig.~\ref{subfig:oran_architecture}.
The model management process involves packaging the model along with the necessary artifacts.
Models are submitted to either the model store or the catalog platform, where they are stored along with the relevant metadata information.
For online training and retraining purposes, other AI/ML components should be able to access these models.
In certain situations, it is possible to upload the entire container image of the model application to the model catalog platform.
The model management component, in coordination with the model inference component and aligned with the requirements of AI/ML assisted solutions, selects the appropriate ML model for deployment.
The model management component has the capability to reselect the model based on feedback and terminate the model in scenarios where severe degradation of model performance is observed~\cite{b1}.

Implementation of model management in the O-RAN Software Community AI/ML Framework project differs depending on the version.
In the latest release, they provide Model Store SDK for storing models and Model Management Service for managing models.
Model Management has a REST (Representational State Transfer) interface and the ability to register models and retrieve a trained model by model ID.
Model Store SDK ensures that models are available as zip archives and exposes them through the URL address~\cite{ai-ml-framework}.
In older releases, AcumosAI was used as a model catalog platform~\cite{Lee2020HostingPlatform}.

Model management is in MLOps terminology referred to as a model registry or model store, as depicted in Fig.~\ref{subfig:mlops_pipeline}.
It is a component that stores, reverts, versions, and manages AI/ML models, while the model deployment phase itself is a separate component in MLOps.
In general model management supports retrieving the latest models by name, ID, or tag.
It has an integrated model storage solution or uses storage software to store entire models with contextual metadata information for model deployment.
Models are stored in a model registry, they can be stored as a file or packaged as an application.
Some model management solutions support both model storage and model deployment.
In this case, a container image of an application is stored, versioned, and then deployed to an inference host by the same software~\cite{Ruf2021DemystifyingTools,Google2020MLOps:Learning}.

Storing a model in a registry is a process of saving model weights in a file with any accessory contextual information needed to load the model again.
Libraries like PyTorch and TensorFlow provide functions for storing model weights, which are then stored as a file.
Alternatively, the model, together with any code needed for loading the model, is packaged into a container image, and the image is stored and versioned by the model management software.

Storage location can be a local, remote file system, or a storage bucket.
A container image registry can also be used for storing and versioning packaged models.

Querying the models can be done directly by referencing the ID, version, and tag, or the latest model can be fetched by the name of the stored model.
The process of fetching the models needs to be available through API to support fetching the latest model by other components~\cite{zaharia2018accelerating}.

\subsection{AI/ML Model Deployment and Inference}
\label{subsec:components:inference}

\begin{figure}[htbp]
\centerline{\includegraphics[width=0.6\linewidth]{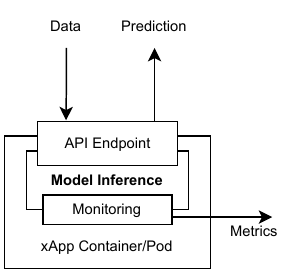}}
\caption{AI/ML model inference on O-RAN.}
\label{fig:inference}
\end{figure}

In the AI/ML inference component, crucial parts are the model inference engine, model deployment phase, and ML model performance monitoring, as defined in O-RAN's specifications.
The inference component integrates with an AI/ML assisted solution and receives model from the model management component, as depicted in AI/ML workflow architecture~\ref{subfig:oran_architecture}.
Deployment with an image or file has to be considered in this component.
In image-based deployments, the AI/ML model is deployed as an xApp (Extended Application), a software application designed to implement network functions or services within the Near-RT RIC~\cite{Orhan2021ConnectionApproach}. The ML inference process runs inside the deployed container, as depicted in Fig.~\ref{fig:inference}.
Images are prepared in such a way that a running container is capable of model inference through an API (Application Programming Interface) endpoint.
For file-based deployments, the AI/ML model is decoupled from the xApp, and the model is deployed on the inference platform running in the xApp.
The image can include the configuration of the inference host and a model description file.
The model inference component can report the performance metrics of the model, such as accuracy, running time, and network KPI (Key Performance Indicator) so that reselection or retraining of the model can be triggered by the AI/ML continuous operation component~\cite{b1, Polese2023UnderstandingChallenges}.

For model deployment and inference, O-RAN SC uses KServe as a model inference platform on Kubernetes.
They developed a KServe adapter which, using a custom .yaml configuration file, deploys stored models.
Their solution allows defining resources for the deployed application, which runs as a Kubernetes pod.
The solution can be integrated with O-RAN SC implementation of Near-RT RIC for the model to be deployed directly as a xApp~\cite{o-ran-sc-docs, ai-ml-framework}.

In MLOps terminology, for model inference component is referred to as model deployment, as depicted in Fig.~\ref{subfig:mlops_pipeline}.
For this purpose, model serving tools are used that manage model deployment and inference procedures.
ML models need to be deployed on an inference host, which can be an edge device or a server in general MLOps systems.
For this purpose, a type of REST API endpoint is needed for accessing the model's prediction service, commonly as a containerized application, which is similar to O-RAN's xApp model inference discussed in the previous paragraph and shown in Fig~\ref{fig:inference}.
This model deployment technique is referred to as a \textit{Model as a Service} deployment~\cite{Ruf2021DemystifyingTools}.
Tools for model deployment can support model inference, the alternative is to build custom container images that can be configured as API endpoints.
Performance metrics need to be collected from the deployed model to support the continuous operation of the AI/ML workflow.
This can be done by separate monitoring software or can be a part of model deployment solution~\cite{zaharia2018accelerating}.

The API endpoint for the deployed model is an interface for the prediction service, where data for model inference is sent as a POST request and prediction data is returned.
This way of accessing the prediction service is the same for MLOps systems and in O-RAN's AI/ML workflow architecture, as depicted in Fig.~\ref{fig:inference}.
The API endpoint is usually accessible as a URL address consisting of an IP address and a port number.
Inference on the API endpoint can be as a REST HTTP or gRPC protocol.
gRPC is a RPC (Remote Procedure Call) framework and offers better performance compared to traditional REST APIs~\cite{brown2023measuring}.
In the case of a distributed system, a load balancing service can be implemented to connect the API endpoint to services on physical machines~\cite{Nguyen2020TowardClusters}.


The inference engine is realized by the model trained in Section~\ref{subsec:components:training} and selected for deployment by the model management discussed in Section~\ref{subsec:components:model_management}.
The performance of the model inference depends on which software and function is chosen.

\subsection{AI/ML Continuous Operation} 
\label{subsec:components:cont_operation}

Machine learning continuous operation functions are not precisely defined, but the specification describes them as a series of functionalities for the continuous improvement of AI/ML models within the entire AI/ML life cycle.
In O-RAN AI/ML workflow architecture they are connected to other components, such as model management and inference, as depicted in Fig.~\ref{subfig:oran_architecture}.
The continuous operation component ensures the operation of all other AI/ML components by verification and monitoring of the workflow and AI/ML model performance.
It collects feedback on model performance and monitoring metrics of all components and triggers corresponding events such as retraining, termination, model reselection, and redeployment.
According to the analysis of the metrics, it recommends and optimizes the components' functions and AI/ML model.
It detects any issues and helps with updating the models without service interruption~\cite{b1, Polese2023UnderstandingChallenges}.

O-RAN Software Community supports AI/ML continuous operation through integration with Kubeflow for their solution.
They developed Kubeflow Adapter, which supports connectivity to Kubeflow.
The Kubeflow Adapter allows the creation of model training pipelines (workflows) and execution of training jobs along with fetching information about pipelines and training jobs.
In addition, the Software Community is developing monitoring software for their solution by using a monitoring server and agent to collect metrics from workflow components and store them in InfluxDB.
The information, metrics, and control over workflows and jobs are available through their Dashboard Portal~\cite{ai-ml-framework}.

AI/ML continuous operation component is related to MLOps pipeline automation and orchestration as well as monitoring.
In Fig.~\ref{subfig:mlops_pipeline} it encompasses all other components, it orchestrates and monitors them.
Note that we use the terms pipeline and workflow interchangeably in this context. 
Depending on the solution, continuous operation can ensure the complete automation of the MLOps workflow or even perform some functions of other components.
The automation of the workflow and the collection of metrics to trigger different tasks (likewise steps, stages, components in different terminologies and software) in a workflow are necessary functionalities of continuous operation.
Moreover, pipeline versioning, artifact storage and caching, and scheduled runs are common functionalities of MLOps pipeline orchestrators that relate to the continuous operation component in the AI/ML workflow of O-RAN~\cite{Google2020MLOps:Learning}.




Monitoring can be a part of every component of an AI/ML workflow system.
In a full AI/ML workflow system, metrics are collected from all the components and presented in a continuous operation or monitoring tool.
Key performance metrics, model drift information, and monitoring of the deployment should be reported or stored for continuous operation to control the workflow and trigger model retraining if necessary~\cite{Kreuzberger2023MachineArchitecture}.

\begin{table*}[ht]
\caption{Types of Data Storage}
\ra{1.2}
\centering
\fontsize{8pt}{10pt}\selectfont 
\begin{tabularx}{0.95\linewidth}{>{\centering\arraybackslash}l|*{4}{>{\centering\arraybackslash}X}@{}}
    \toprule
    \textbf{\textit{Type}} &
    \thead{\textbf{\textit{Storage Bucket}}} &
    \thead{\textbf{\textit{Data Lake}}} &
    \thead{\textbf{\textit{Data Warehouse}}} &
    \thead{\textbf{\textit{Database}}} \\ \midrule\midrule
    
    \textit{Tools} &
    MinIO, Ceph, Apache Ozone &
    Apache (Hudi, Iceberg, Flink, Kafka, Hadoop), Kylo, Delta Lake &
    Apache (Hive, Doris, Kylin) &
    InfluxDB, Apache (Cassandra, HBase), MongoDB, PostgreSQL \\ \midrule
    
   \textit{Data Retrieval} &Fetching by name& SQL Queries&SQL Queries&SQL Queries \\ \midrule
    
    \textit{ETL} &\ding{55}&\ding{51}&\ding{51}&\ding{55} \\ \midrule

    \textit{Data Types} &Unstructured&Any&Structured&Structured \\ \midrule

    
\end{tabularx}
\label{tab:storage}
\end{table*}

\section{Tool Analysis} \label{sec:tools}
In this section we compare open-source tools for each of the AI/ML components based on general MLOps definitions and O-RAN AI/ML workflow architecture analysis presented in Section~\ref{sec:workflow_architecture}.
This analysis identifies the strengths and limitations of each tool, providing insights into their suitability for different AI/ML tasks. 
The comparison of features and analysis of the tools serve as a guideline to select the appropriate tools according to specific project requirements and to simplify the process of creating an AI/ML and MLOps systems.

The analysis and comparison of these tools is based on a thorough review of cited official documentation and research papers by authors associated with tool development or related research papers utilizing the analyzed tools and these sources are appropriately cited.
Hands-on experimentation was conducted with selected tools, specifically with all the tools used by our proposed solution in Section~\ref{sec:solution} evaluated in Section~\ref{sec:results} to confirm the conclusions derived from this analysis. 

\subsection{Open-Source Data Storage}
\label{subsec:tools:data_storage}
For storing data in the various stages of the ETL process as discussed in Section~\ref{subsec:components:data_prep} and depicted in Fig.~\ref{subfig:oran_architecture}, various types of data storage solutions exist. A large number of the open-source solutions are provided by the Apache Software Foundation, but also other open source solutions exist as can be seen from some of the tools mentioned in the first row of Table~\ref{tab:storage} while commercial and cloud options such as Amazon Redshift, and Google BigQuery that offer a complete managed storage and ETL solution~\cite{Nambiar2022AnManagement} are beyond the scope of this paper due to the openness requirement in Section~\ref{sec:requirements}.
To realize a storage system with open-source software multiple tools, falling into four broad categories as per the columns on Table~\ref{tab:storage}, are needed.

\paragraph*{Storage Bucket}
Storage buckets are object storage tools for storing unstructured data in flat structures grouped into buckets as presented in the first column of Table~\ref{tab:storage}.
Open-source storage buckets are MinIO, which is an object store for AI data infrastructure~\cite{minio}, Ceph, a distributed storage system with an object store service~\cite{cephdocs} and Apache Ozone, a highly scalable and distributed storage~\cite{apacheozone}.

\paragraph*{Data Lake}
Data lakes are repositories for storing raw data in their native format~\cite{datalakes} and include tools such as Apache Hudi, Kylo, and Delta Lake as summarized in the third column of Table~\ref{tab:storage}.
Apache Hudi can be used as a data lake~\cite{apachehudi}.
Apache Kafka is used for streaming data, data ingestion, and data pipelines~\cite{apachekafka}.
Apache Hadoop and Flink are data processing engines for Big Data~\cite{apachehadoop, apacheflink}.
Some or more of the Apache tools listed can be used to realize the full functionality of a data lake.
Kylo and Delta Lake are complete data lake solutions~\cite{kyliowebsite, deltaio}.

\paragraph*{Data Warehouses}
Data warehouses are similar to data lakes but for structured data~\cite{Nambiar2022AnManagement} and include Apache tools such as Hive, Doris, and Kylin as presented in the forth column of Table~\ref{tab:storage}.
Apache Hive is a distributed data warehouse system~\cite{hivewebsite}.
Apache Doris is a data warehouse for real-time analytics~\cite{doriswebsite}.
Apache Kylin is a distributed analytical data warehouse for Big Data~\cite{kylinwebsite}.

\paragraph*{Database}
Databases are high-performant storage for structured data and utilize query language for fetching the data, as summarized in the last column of Table~\ref{tab:storage}.
Apache Cassandra and MongoDB are NoSQL databases~\cite{lakshman2010cassandra,mongodbwebsite}.
Apache HBase is a Hadoop database for Big Data~\cite{hbasewebsite}.
InfluxDB is a performant time series database service~\cite{influxdatawebsite}.
PostgreSQL is a relational database management system~\cite{postgresqlwebsite}.

\subsubsection{Analysis}
Different types of data storage, presented in Table~\ref{tab:storage}, can be utilized depending on the use case and requirements.
Data lakes support data processing and querying.
Data lakes are often utilized in production environments and can support the entire ETL process.
Data warehouses are similar to data lakes but for structured data.
Data warehouses and data lakes usually utilize underlying storage systems such as storage buckets or file systems.
Most open-source tools for data warehouses and data lakes are from the Apache Software Foundation.
Those tools integrate well with each other and a combination of those tools can be used to realize an entire data lake solution.

Storage buckets are object/blob storage solutions.
In contrast to normal file systems, they store data in flat structures that are grouped into buckets.
The data is retrieved by fetching by name.
ETL processes or SQL queries are not supported.
Compared storage bucket software differs in complexity and maturity.
All three support storing objects over S3 protocol.
Red Hat Ceph is more complex to deploy than MinIO, both of them can be deployed on Kubernetes.
Apache Ozone is less mature but offers Hadoop-compatible file system implementation and it integrates well with other Apache tools.

Databases are structured storage solutions with low latency for querying data.
Some of the solutions such as InfluxDB offer data processing functions.
Apache databases integrate better with the Apache ecosystem and are made for Big Data.
PostgreSQL is a more traditional SQL database management system.

The decision on which data storage type and tool to select for the project depends on the performance requirements, data retrieval and processing needs, and integration with other processing tools and AI/ML training tools.

\begin{table*}[ht]
    \caption{Data Preparation Tools}
    \ra{1.2}
    \centering
    \fontsize{8pt}{10pt}\selectfont 
    \begin{tabularx}{0.7\linewidth}{@{}>{\centering\arraybackslash}X|*{6}{>{\centering\arraybackslash}p{1.2cm}}@{}}
        \toprule
        \thead{\textit{\textbf{Features\textbackslash Tools}}} & 
        \textit{\textbf{Pachyderm}} & 
        \textit{\textbf{Feast}} & 
        \textit{\textbf{Spark}} & 
        \textit{\textbf{GX}} & 
        \textit{\textbf{Ray Data}} & 
        \textit{\textbf{DVC}} \\
        \midrule\midrule
        Extraction & \ding{51} & \ding{51} & \ding{51} &  & \ding{51} & \ding{51} \\
        \midrule
        Transformation &\ding{51}  & \ding{51} & \ding{51} & \ding{51} & \ding{51} & \\
        \midrule
        Validation & \ding{51} & \ding{51} & \ding{51} & \ding{51} &  & \\
        \midrule
        Train/Test Split & \ding{51} &  & \ding{51} & \ding{51} & \ding{51} & \ding{51} \\
        \midrule
        Feature Store &  & \ding{51} &  &  &  &  \\
        \midrule
        Versioning & \ding{51} &  &  &  &  & \ding{51} \\
        \bottomrule
    \end{tabularx}
    \label{tab:prep}
\end{table*}

\subsection{Open-Source Data Preparation Tools Analysis}
\label{subsec:tools:data_prep}

In this subsection open-source data preparation tools are analyzed and compared.
Tools are compared by features they support based on O-RAN Alliance specifications and general MLOps requirements, as discussed in Section~\ref{subsec:components:data_prep}. 
These tools solve the requirements of the data preparation component in AI/ML workflow architecture depicted in Fig.~\ref{subfig:oran_architecture}.
All the compared tools are open-source and can be installed on on-premise infrastructure.

The following open-source tools were chosen and compared for the data preparation component and the comparison of features they support is presented in Table~\ref{tab:prep}.

DVC is a data management and versioning tool with support for data and ML pipelines~\cite{dvc}.
Feast is a standalone feature store for storing and serving features~\cite{feast}. 
In contrast, Great Expectations is a Python framework for profiling and validating the state of data~\cite{gx}.
As an end-to-end data preparation tool, Pachyderm Community Edition offers a solution for data pipeline automation, data versioning, and tracking data lineage~\cite{pachyderm}.
Ray is a compute framework for scaling AI and Python applications and supports distributed data transformation functionalities and data loading through its Ray Data library ~\cite{ray}.
Apache Spark is a data processing engine for data science~\cite{Zaharia2016ApacheProcessing}.

\subsubsection{Analysis}
These tools support different requirements listed in the first column of Table~\ref{tab:prep}.
Apache offers a range of tools for data storage, processing, and data science.
For data preparation, Apache Spark and PySpark provide low-level control over data transformation and validation.
Besides data preparation, Spark offers a solution for AI/ML training component.
Similarly, Ray Data integrates well with Ray's Train library, but data preparation is not a primary function of Ray.
Both Spark and Ray Data offer distributed batch data transformations and parallel computing.

DVC primarily serves as a data versioning tool and supports the creation of data processing and ML pipelines.
Together with Git, it provides a solution for development and collaboration on data science projects.
Feast and Great Expectations OSS support transformation and validation, with Feast providing feature engineering and feature store, while Great Expectations OSS is a lightweight library for validating data structures.

Combining a selection of these tools with a storage solution enables the creation of a data pipeline tailored to the specific requirements of a project.
Pachyderm, on the other hand, offers a comprehensive end-to-end solution with a user interface for data preparation, but it's worth noting that only the Community Edition is free and open-source.
Pachyderm requires Kubernetes to operate and can be more challenging to set up on an on-premise infrastructure than other tools reviewed.

\subsection{Open-Source Model Training Tools}
\label{subsec:tools:model_training}

\begin{table}[htbp]
\caption{Model Training Tools}
\ra{1.2}
\centering
\fontsize{8pt}{10pt}\selectfont 
\begin{tabularx}{\linewidth}{@{}>{\centering\arraybackslash}X|*{4}{>{\centering\arraybackslash}p{1.1cm}}@{}}
\toprule
\thead{\textit{\textbf{Features\textbackslash Tools}}} & \textit{\textbf{Dask}} & \textit{\textbf{Spark}} & \textit{\textbf{Ray}} & \textit{\textbf{Optuna}}  \\
\midrule\midrule
Hyperparameter Tuning & \ding{51} & \ding{51} & \ding{51} & \ding{51} \\
\midrule
Evaluation &  & \ding{51} & &  \\
\midrule
Experiment Tracking & \ding{51} & \ding{51} & & \ding{51} \\
\midrule
Distributed Training & \ding{51} & & \ding{51} &  \\
\midrule
Dashboard & \ding{51} & \ding{51} & \ding{51} & \ding{51} \\
\midrule
Multi Library Support & \ding{51} & & \ding{51} & \ding{51} \\
\bottomrule
\end{tabularx}
\label{tab:train}
\end{table}

Various frameworks and libraries have been developed to train AI/ML models.
Some provide control over model design, selection, and training parameters and some are high-level platforms for automated and distributed training.
In this analysis, we compare high-level frameworks that provide ML training management and functionalities from MLOps and AI/ML workflow viewpoints.
The comparisons of tools for model training, aligned with O-RAN's requirements and MLOps parameters discussed in Section~\ref{subsec:components:training}, are provided in Table~\ref{tab:train}.

Ray is a compute framework supporting diverse AI/ML distributed training functionalities through its libraries: Ray Core, Ray Train, and Ray Tune\cite{Moritz2007Ray:Applicationsb, ray}.
Dask is a Python library for parallel and distributed computing.
Dask-ML provides scalable machine learning using common ML library integrations~\cite{daskml}.
Similarly, Apache Spark is a unified engine for data science with MLlib, which is its library for machine learning~\cite{Assefi2017BigMLlib}.
On the other hand, Optuna is hyperparameter optimization software~\cite{optuna_2019}.

\begin{table*}[htbp]
\caption{Tools with Model Management Functionalities}
\ra{1.2}
\centering
\fontsize{8pt}{10pt}\selectfont 
\begin{tabularx}{0.8\linewidth}{@{}>{\centering\arraybackslash}X|*{6}{>{\centering\arraybackslash}p{1.5cm}}@{}}
\toprule
\thead{\textit{\textbf{Features\textbackslash Tools}}} & \textit{\textbf{MLflow}} & \textit{\textbf{BentoML}} & \textit{\textbf{Modelstore}} & \textit{\textbf{ZenML}} & \textit{\textbf{AcumosAI}} & \textit{\textbf{DVC}}\\
\midrule\midrule
Storage Bucket & \ding{51} & \ding{51} & \ding{51} & \ding{51} &  &  \ding{51}\\
\midrule
Metadata & \ding{51} & \ding{51}  & \ding{51} &  & \ding{51} &  \\
\midrule
Versioning & \ding{51} & \ding{51} & \ding{51} & \ding{51} & \ding{51} & \ding{51} \\
\midrule
Packaging & \ding{51} &\ding{51}   &  &  & \ding{51} &  \\
\midrule
Dashboard & \ding{51} &  &  & \ding{51} &  &  \\
\bottomrule
\end{tabularx}
\label{tab:model-components}
\end{table*}

\subsubsection{Analysis}
There are several libraries and frameworks available for training ML models.
When comparing frameworks with more high-level functionalities and distributed learning capabilities, we limit ourselves to Ray, Dask, and Apache Spark.
The main difference is that Ray emphasizes machine learning compared to the other two tools and supports a wider range of ML features through its libraries.
Both Ray and Dask can utilize Python libraries such as PyTorch and TensorFlow to train models, while Spark relies on its own MLlib library.
Dask excels in supporting the tracking and logging of experiments, whereas Ray fulfills more requirements of other AI/ML components such as model serving and data processing.

Optuna stands out as a dedicated hyperparameter optimization tool that can work independently or be integrated with other software for model training.
Unlike the tools with integrated hyperparameter tuning, Optuna supports pruning, has a lightweight structure, and includes a dashboard specifically designed to monitor the progress of hyperparameter tuning.

\subsection{Open-Source Model Management Tools}
\label{subsec:tools:model_management}
Open-source tools for model management can be a simple model storage solution or a comprehensive platform for managing, versioning, and deploying AI/ML models.
Certain machine learning platforms integrate a model store within their artifact storage, alongside other AI/ML functionalities.
Specific tools have been developed exclusively to manage and store models.
According to the mentioned features in the preceding subsections, six open-source model management tools were examined and the results are presented in Table~\ref{tab:model-components} and discussed in the following paragraphs.

MLFlow is an open-source experiment tracker and model packaging MLOps tool. One of its components, MLFlow Models, is suitable for packaging code and data for AI/ML models that can work with various deployment tools~\cite{zaharia2018accelerating}.
ZenML, a sophisticated framework for automating machine learning pipelines, includes ML pipeline components. Among these is an artifact store that can function as a simple model store~\cite{zenml}.
Modelstore offers a lightweight Python library that provides a solution for machine learning model registry~\cite{modelstore}.
AcumosAI serves as an AI platform and marketplace designed for continuous learning of AI models. This platform can manage, store, and package models for deployment~\cite{Zhao2018PackagingPlatform}.
Lastly, BentoML is an end-to-end platform for managing and deploying machine learning models, featuring a built-in model store and packaging solutions~\cite{bentoml}.

\begin{table*}[htbp]
\caption{Tools for Model Inference and Deployment}
\ra{1.2}
\centering
\fontsize{8pt}{10pt}\selectfont 
\begin{tabularx}{0.85\linewidth}{@{}>{\centering\arraybackslash}X|*{7}{>{\centering\arraybackslash}p{1.4cm}}@{}}
\toprule
\thead{\textit{\textbf{Features\textbackslash Tools}}} & \textit{\textbf{Ray Serve}} & \textit{\textbf{KServe}} & \textit{\textbf{MLflow}}& \textit{\textbf{BentoML}} & \textit{\textbf{TFServing}} & \textit{\textbf{TorchServe}}  & \textit{\textbf{FastAPI}}\\
\midrule\midrule
Distributed Serving & \ding{51} & \ding{51}  & &  &  &  &  \\
\midrule
Support for gRPC & \ding{51}  &\ding{51}  &   & \ding{51} & \ding{51} & \ding{51} &  \\
\midrule
Dashboard &  \ding{51} & \ding{51} & \ding{51} & & & & \\
\midrule
Multi Library Support & \ding{51} & \ding{51} & \ding{51}& \ding{51} & &  &\ding{51}  \\
\bottomrule
\end{tabularx}
\label{tab:inference}
\end{table*}

\subsubsection{Analysis}
Model management solutions differ in the functionalities they support and whether they offer solutions for model packaging and deployment.
Lightweight solutions such as Modelstore and ZenML's artifact store have only basic features like versioning and storage.
While AcumosAI and BentoML platforms offer packaging models into end container applications ready for deployment.
DVC offers basic versioning functionality but is not primarily meant for storing models.
MLflow is a more lightweight solution compared to platforms such as AcumosAI and BentoML, while still offering model management features like model metadata storage and dashboard.

All the compared tools except AcumosAI support model storage in a storage bucket location, which is important when fetching models in a distributed system and from a remote/edge machine.
As model management is highly coupled with model deployment and inference, the choice depends on the requirements and tools chosen for the AI/ML Inference component, whether models need to be packaged by the model management software, as well as the number and type of models.

\subsection{Open-Source Tools for Model Deployment and Inference}
\label{subsec:tools:model_inference}

In this subsection open-source tools for model O-RAN model inference component, discussed in Section~\ref{subsec:components:inference},  are listed and compared.
Certain solutions are part of the platform for model management while some tools are integrated with machine learning training libraries.
Tools differ in whether they serve models as a packaged application or only provide API endpoint for model inference.
Open-source tools that support model deploying and inference are presented in Table~\ref{tab:inference}.

MLflow Models supports deploying models or creating containerized applications ready for inference~\cite{mlflow}. 
Similarly, BentoML supports model deployment or creation of packaged images with model endpoints~\cite{bentoml}.
Ray Serve, a Ray library for serving models, supports inference through FastAPI or ML libraries~\cite{ray}. 
KServe, a model inference platform, provides a Kubernetes CRD (Custom Resource Definition) for serving machine learning models on arbitrary frameworks~\cite{kserve}.
TensorFlow Serving is a model serving system specifically designed for performant serving of TensorFlow models~\cite{olston2017tensorflowserving}, while TorchServe is a tool dedicated to serving PyTorch models~\cite{torchserve}.
Lastly, FastAPI, a web framework for building APIs, can be used in conjunction with ML libraries for serving ML models~\cite{fastapi}.

\subsubsection{Analysis}
Model management tools like MLflow and BentoML support model deployment out of the box.
They work best if used as a model management tool in the AI/ML workflow.

ML training libraries such as TensorFlow's TFServing and PyTorch's TorchServe often support model serving, which can simplify the workflow serving the models in the same step as training.
These libraries increase inference performance for TensorFlow and PyTorch, by using server-side batching and other optimizations for higher throughput~\cite{mux_tensorflow_serving}.
All examined tools except MLflow and FastAPI support gRPC endpoints, which offer better performance compared to REST APIs, as explained in Section~\ref{subsec:components:inference}. The process of setting up this service can be more complicated than REST API endpoints.

Ray Serve and KServe have an advantage in supporting model serving on a distributed Kubernetes cluster, but KServe is designed to work with the Kubeflow orchestrator.
Kubernetes deployed models can utilize supplementary services for high availability, horizontal scaling, and load balancing for larger model inference requirements.
KServe serves models using lower-level serving libraries like TFServing and TorchServe to get the benefits of increased performance for inference, while other high-level model serving frameworks usually utilize libraries' predict function for inference.
Frameworks that allow model deployment on Kubernetes can be harder to set up than tools like MLflow and BentoML.

\begin{table*}[htbp]
\caption{Workflow Orchestration, Continuous Operation and Monitoring Tools}
\ra{1.2}
\centering
\fontsize{8pt}{10pt}\selectfont 
\begin{tabularx}{1\linewidth}{@{}>{\centering\arraybackslash}X|*{8}{>{\centering\arraybackslash}p{1.4cm}}@{}}
\toprule
\thead{\textit{\textbf{Features\textbackslash Tools}}} & \textit{\textbf{Flyte}} & \textit{\textbf{ZenML}} & \textit{\textbf{Kubeflow}} & \textit{\textbf{Airflow}} & \textit{\textbf{Metaflow}} & \textit{\textbf{DVC}}  & \textit{\textbf{Prometheus / Grafana}} & \textit{\textbf{Deepchecks}}  \\
\midrule\midrule
Scheduled Runs & \ding{51} & \ding{51} & \ding{51} & \ding{51} &  &  &  &  \\
\midrule
Artifact Store & \ding{51} & \ding{51} & \ding{51} & \ding{51} & \ding{51} & \ding{51} &  &  \\
\midrule
Artifact Caching & \ding{51} & \ding{51} & \ding{51} &  &  & \ding{51} &  &  \\
\midrule
Containerized Steps & \ding{51} & \ding{51} & \ding{51} & \ding{51} & \ding{51} &  &  &\\
\midrule
Dashboard & \ding{51} & \ding{51} & \ding{51} & \ding{51} & \ding{51} &  & \ding{51} & \ding{51} \\
\midrule
Kubernetes & \ding{51} & \ding{51}* & \ding{51} & \ding{51}* & \ding{51}* &  & \ding{51}* &  \\
\midrule
Workflow Versioning & \ding{51} & \ding{51} &  & \ding{51} & \ding{51} & \ding{51} &  &  \\
\midrule
Monitoring &  &  &  &  & \ding{51} &  & \ding{51} & \ding{51} \\
\bottomrule
\multicolumn{9}{l}{*Tool can be deployed on varying back-ends.}
\end{tabularx}
\label{tab:orchestration}
\end{table*}

FastAPI is, compared to other tools, not primarily intended for serving models but for creating APIs.
The advantage is that it is lightweight, but it needs a supplementary server like Uvicorn to deploy the API endpoint \cite{uvicorn}.
It can be used in conjunction with Ray Serve to simplify the process of creating Ray API endpoints or it can be a stand-alone solution with ML training libraries.
BentoML is an end-to-end solution for model management and serving, but it doesn't offer distributed serving and some functionalities are only available with a BentoML cloud account.

\subsection{Open-Source Tools for Continuous Operation and Workflow Orchestration}
\label{subsec:tools:cont_operations}
The comparison of open-source tools for the AI/ML continuous operation component consists of tools primarily focused on workflow automation, orchestration, and monitoring tools.
Workflow automation tools differ in the supported integration with other open-source tools and in the simplicity of development and deployment.
Table~\ref{tab:orchestration} presents these tools for orchestrating machine learning workflows and pipelines, tools for monitoring, and the features these tools provide.

ZenML is a pipeline automation framework featuring interchangeable back-end components for orchestrator, artifact store, and model deployment software~\cite{zenml}. Similarly, Flyte offers workflow automation on top of Kubernetes~\cite{flytedocs}.
Kubeflow provides a collection of components for orchestrating and developing ML workloads on Kubernetes~\cite{kubeflow}, while Airflow serves as a general task and workflow orchestrator~\cite{airflow}.
DVC enables pipeline definitions in .yaml files for data and ML pipelines~\cite{dvc}, and Metaflow functions as a workflow orchestrator specifically for AI/ML projects~\cite{metaflow}.
For continuous evaluation of AI models and data, Deepchecks offers a dedicated tool~\cite{Chorev_Deepchecks_A_Library_2022}. 
In the realm of monitoring, Prometheus handles scraping and collecting metrics, while Grafana provides a dashboard to visualize these metrics, both applicable to AI/ML metrics~\cite{prometheus, grafana}.

\subsubsection{Analysis}
When evaluating open-source tools for the continuous operation component and workflow automation for the AI/ML workflow system, features to consider are the specific functionalities, ease of use, and deployment characteristics of each tool as these are the main differences between analyzed pipeline automation tools.

ZenML stands out as a flexible pipeline automation framework with the unique feature of interchangeable back-end components for orchestrator, artifact store, experiment tracking, and model deployment. 
This flexibility provides users with the ability to change their setup according to specific project requirements.
It allows changing from local to containerized execution of steps and rapid switching between development and production environments.

Flyte, which is a workflow automation framework built on Kubernetes, compared to ZenML offers advantages in terms of scalability as it can scale to thousands of workflows but is not as flexible in terms of back-end components.
Registering and running registered workflows in a production environment is supported in Flyte.
Meanwhile, Kubeflow is a collection of components for ML workload orchestration on Kubernetes and provides extensive support for various AI/ML component requirements. 
However, the trade-off lies in the complexity of its setup. Both Kubeflow and Flyte require Kubernetes for operating in production which can be challenging to set up, while ZenML offers varying deployment back-ends. 

Airflow stands out as a general-purpose workflow orchestrator and pairs well with other Apache tools for data processing and machine learning.
While Metaflow is a workflow orchestrator designed specifically for AI/ML projects and has a balance between simplicity and functionality.
DVC, while offering pipeline definitions for data and ML pipelines, is better suited for simpler tasks and development setup, lacking the advanced orchestration, dashboard, and workflow scheduling capabilities found in other analyzed tools.

On the monitoring side of AI/ML continuous operation, Prometheus as a metric scraper and Grafana for metric visualization can be used in AI and non-AI projects. 
They are designed to collect and present any type of metrics and are highly customizable.
They integrate with AI/ML tools by scraping exported metrics of those tools.
The alternative is Deepchecks, which is a sophisticated evaluation tool for monitoring and evaluating deployed ML models along with monitoring data for training and inference.

\section{The Proposed NAOMI Solution} 
\label{sec:solution}

\begin{figure}[htbp] 
\centerline{\includegraphics[width=1\linewidth]{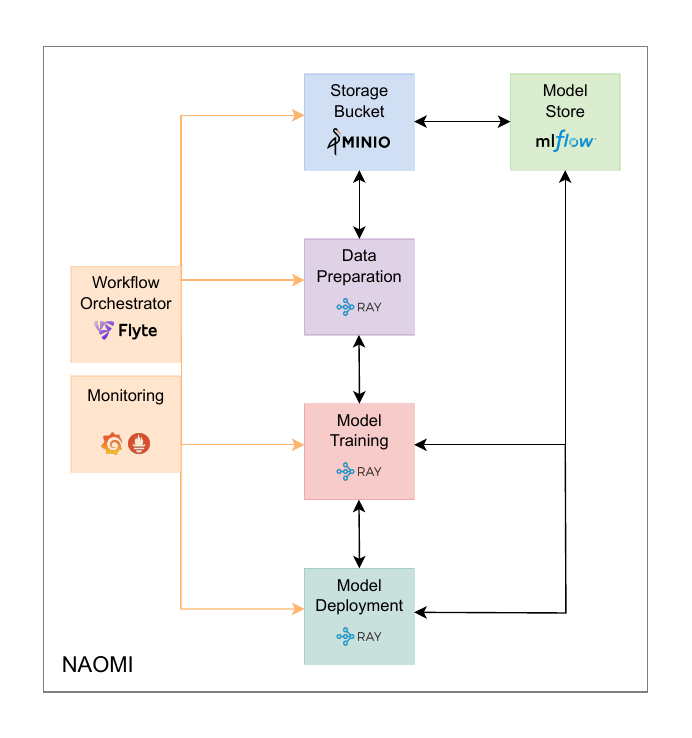}}
\caption{Architecture diagram of the proposed NAOMI - AI/ML workflow democratized solution.}
\label{fig:nancy-arch}
\end{figure}

Guided by the requirements for democratization from Section~\ref{sec:requirements}, architecture analysis in Section~\ref{sec:workflow_architecture}, and the findings of the tool analysis in Section~\ref{sec:tools} we propose NAOMI, as depicted in Fig.~\ref{fig:nancy-arch}, a new solution that embraces O-RAN principles and workflow architecture illustrated in Fig.~\ref{subfig:oran_architecture}, aligns with MLOps vision and enables democratization through careful tool selection.
NAOMI architecture shows a flow of operations and data artifacts inside the AI/ML workflow. 
Further, Fig.~\ref{fig:nancy-arch} color maps components in the proposed architecture to the one from O-RAN, presented in Fig.~\ref{subfig:oran_architecture} and to the general MLOps architecture illustrated in Fig.~\ref{subfig:mlops_pipeline}.

For the data preparation component and data storage solution, analyzed in Section~\ref{subsec:components:data_prep}, we selected MinIO~\cite{minio}, a simple storage bucket solution. 
As discussed in Section~\ref{subsec:tools:data_storage}, for more democratized, non-enterprise environments, storage buckets offer a simpler solution with less overhead compared to alternatives, satisfying the ease of use requirement identified in Section~\ref{sec:requirements}.
To support modularity and ease of use, data transformation is enabled by the Ray framework through its Ray Data library or any other Python library for data transformations such as Pandas, providing a modular approach without locking the user to a single platform and simplifying the integration of existing ML code.

Similarly, for the model training component, analyzed in Section~\ref{subsec:components:training}, we selected the Ray framework and its ML libraries. According to the analysis Section~\ref{subsec:tools:model_training}.
Ray supports a modular approach with little to no modification to run existing machine learning training procedures. 
Users can choose the preferred model training library such as TensorFlow, Sklearn, and PyTorch, which can be integrated into Ray and adapted for distributed model training. 
Further, Ray libraries enable distributed training on multiple architectures, which support the heterogeneity requirement identified in Section~\ref{sec:requirements}. 

The model management component is analyzed in Section~\ref{subsec:components:model_management} and
MLflow is utilized as a model management component, as it supports most of the analyzed features discussed in Section~\ref{subsec:tools:model_management} and summarized in Table~\ref{tab:model-components}.
The training component in the workflow can fetch models stored in MLflow for model retraining and store the updated models in MLflow.
It can act as a separate component utilized by actors outside the AI/ML workflow cluster to provide access to the latest models acting as an independent component and supporting the modularity requirement identified in Section~\ref{sec:requirements}. 

The model inference component, described in Section~\ref{subsec:components:inference}, is enabled by Ray Serve.
Providing distributed serving and integration with other Ray libraries for model training, as explained in Section~\ref{subsec:tools:model_inference}, it enables distributed model inference without needing to rely on additional model deployment and inference tools. This approach simplifies the architecture and through Ray hardware architecture independence supports heterogeneity, ease of use, and distributed services requirements identified in Section~\ref{sec:requirements}.

The continuous operation component, analyzed in Section~\ref{subsec:components:cont_operation}, is separated into workflow orchestration and monitoring components in our proposed architecture, depicted in Fig.~\ref{fig:nancy-arch}. 
The system utilizes Flyte as a workflow orchestrator as it offers advantages in terms of scalability compared to alternatives, as analyzed in Section~\ref{subsec:tools:cont_operations} and supports the scalability requirement of a democratized solution.
Prometheus and Grafana are utilized for monitoring, as they have support for monitoring Kubernetes clusters as well as other tools deployed on Kubernetes according to the analysis in Section~\ref{subsec:tools:cont_operations}.

NAOMI is deployed on Kubernetes, a system for automated deployment, scaling, and orchestration of containerized applications~\cite{Kubernetes}.
Kubernetes satisfies requirements, such as virtualization, distributed services, and scalability of the system, as identified in Section~\ref{sec:requirements}.
Further, NAOMI can be deployed on a heterogeneous Kubernetes cluster comprising multiple ARM and x86\_64 GNU/Linux nodes to accommodate the heterogeneity of network edges. 
This deployment scenario mimics the deployment on distributed edge infrastructure, RAN's network slices~\cite{Afolabi2018NetworkSolutions}, and MEC (multi-access edge computing) devices, which offer cloud computing capabilities at the RAN edge near end users~\cite{Taleb2017OnOrchestration}.
NAOMI can utilize the computational capabilities of the entire distributed cluster for different components of the AI/ML workflow.

\subsection{Support for Democratized System Requirements}
\label{subsec:support_for_dem}

The requirements for a democratized AI/ML workflow solution are identified in Section~\ref{sec:requirements}.
Table~\ref{tab:req} lists which requirements NAOMI supports comparing it to the O-RAN AI/ML Framework solution introduced in Section~\ref{sec:workflow_architecture}.
NAOMI satisfies all listed requirements, while the O-RAN AI/ML Framework supports four out of eight requirements.

\begin{table}[htbp]
\caption{Comparison of Supported Requirements}
\ra{1.2}
\centering
\fontsize{8pt}{10pt}\selectfont 
\begin{tabularx}{1\linewidth}{@{}>{\centering\arraybackslash}X|*{2}{>{\centering\arraybackslash}p{1.8cm}}@{}}
\toprule
\thead{\textit{\textbf{Requirements}}} & \textit{\textbf{NAOMI}} & \textit{\textbf{O-RAN}} \\
\midrule\midrule
 Openness & \ding{51} &  \ding{51} \\
 \midrule
Virtualization & \ding{51} & \ding{51} \\
\midrule
Distributed Services & \ding{51} & \ding{51} \\
\midrule
Scalability & \ding{51} & \ding{51} \\
\midrule
Ease of Use & \ding{51} & \\
\midrule
Modularity & \ding{51} &  \\
\midrule
Self-Evolving & \ding{51} &  \\
\midrule
Heterogeneity & \ding{51} &  \\
\bottomrule
\end{tabularx}
\label{tab:req}
\end{table}

Both solutions rely on open-source tools and are open for anyone to deploy, supporting the openness requirement.
They operate on top of Kubernetes with containerized tools deployed as a Kubernetes application, meeting the virtualization and distributed services requirements, though the extent of support for distributed services differs.
This approach makes both solutions scalable as the number of nodes in the Kubernetes cluster increases.




Regarding ease of use, NAOMI focuses on simple deployment and usage procedures with minimal manual commands. According to the identified requirement in Section~\ref{sec:requirements}, any solution should be set-up with a few simple steps. 
While O-RAN installation documentation explains the installation procedure, it requires more than ten steps with multiple commands each to successfully deploy and utilize the solution~\cite{o-ran-sc-docs}. 
On the other hand, NAOMI requires three steps to deploy the solution and three steps to execute example AI/ML workflows on the system.

Similarly, the modularity of the system is central to NAOMI.
Tools are disabled with a simple configuration file and used independently or integrated into a workflow.
O-RAN utilizes open-source tools by developing interfaces that do not allow modification of the tools used for AI/ML components or configuring them separately.

Further, NAOMI supports collecting metrics, monitoring, and automated retraining and redeployment of models.
The identified self-evolving requirement in Section~\ref{sec:requirements} requires automatic retraining and management of models, which, in the O-RAN solution, is a manual process.

NAOMI can be deployed on ARM and x86 devices, supporting heterogeneity in deployment on network edges. In contrast, O-RAN doesn't support hardware architecture independence, as it can't be deployed on ARM architecture.

To support these claims, we perform an architecture analysis, evaluation and discussion of democratization requirements supported by NAOMI but not by O-RAN, namely \textit{ease of use, modularity, self-evolving, and heterogeneity}, in Section~\ref{subsec:eval_requirements}.

\subsection{NAOMI User Workflow}

\begin{figure}[!htb]
\centerline{\includegraphics[width=1\linewidth]{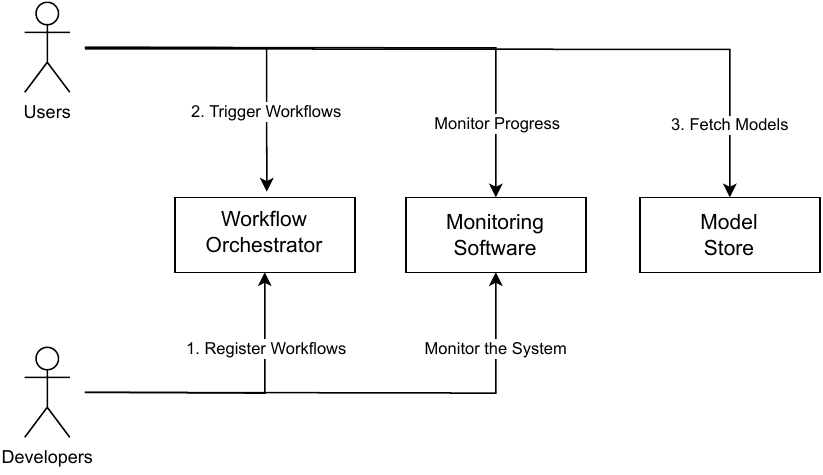}}
\caption{Workflow diagram of actors interacting with the proposed NAOMI AI/ML workflow solution.}
\label{fig:nancy-workflow}
\end{figure}

NAOMI is a production system that runs partially or fully autonomously with users interacting with the system following the user workflow diagram in Fig.~\ref{fig:nancy-workflow}.

Two different types of actors interact with the NAOMI components: developers and users.
In the first step, developers register workflows to the Flyte workflow orchestrator.
Flyte workflows are production AI/ML workflows, which include the AI/ML procedure for transforming the data and training the model.
Flyte workflows can automate the model retraining and redeployment based on monitoring metrics.
Developers have access to dashboards and monitoring software to monitor the status of the deployed solution.

After developers register the workflows, users with the appropriate access to the system trigger the workflows manually or on a scheduled cadence in step two depicted in Fig.~\ref{fig:nancy-workflow}.
They monitor the progress of the machine learning procedure and the status of the syApplicabilitystem.

When one of the AI/ML workflows trains new models, they are stored in the model store and can be accessed by other AI/ML workflows that retrain and redeploy the models or by users, as depicted in step three in Fig.~\ref{fig:nancy-workflow}.

\subsection{General Applicability}
\label{subsec:general_applica}
NAOMI is designed with flexibility in mind applying to a wide range of AI/ML use cases. It also supports diverse model types and workflows. 
NAOMI's advantage is its deployment in large-scale network environments with devices on the network edge or distributed production systems where continuous delivery and model management are important. 
NAOMI's automated workflows simplify the approach to production AI/ML by enabling frequent model retraining, redeployment, and scaling. 
The built-in scheduling, workflow orchestration, and the shared model store enable the process of maintaining models across networks of devices, specifically when a subset of those devices are on the network edge. Therefore, a key strength of NAOMI is its ability to scale across heterogeneous computing environments. It supports model training and inference on x86 and ARM-based devices, making it applicable for setups with edge computing and inference nodes where resources are constrained while enabling the option to offload the computation on central nodes.

In smaller or single-machine setups, this same level of automation can introduce an additional layer of complexity and overhead. 
When running isolated experiments or lightweight testing, the Kubernetes-based approach likely exceeds the operational needs. 
While NAOMI can function in these more constrained environments, a more direct execution approach is more efficient.

More specifically, NAOMI is best suitable for workflows based on commonly used machine learning and data science libraries such as TensorFlow, Keras, and PyTorch. However, virtually any Python based ML code can be ported to work on NAOMI. For example, workflows for Quality of Experience prediction, image recognition, or MNIST handwritten digit classification are all suitable to run on NAOMI, as we show in Section~\ref{subsec:resoults:general_models}.
Further, a potentially wider selection of use cases is enabled by NAOMI. For example, NAOMI can function as a stand-alone model store by enabling only the MLFlow module and, this way, providing an ML model storage solution without the overhead of workflow orchestration or monitoring. Similarly, Ray is a powerful platform on its own, enabling use cases such as distributed ML model training and inference, and even distributed computation use cases beyond AI/ML. 
By enabling only the Ray module, these use cases can be facilitated with lower resource cost than when installing the entire NAOMI system.

\section{Evaluation Methodology} 
\label{sec:methodology}

To evaluate the proposed NAOMI solution we compare it to the current release of the O-RAN AI/ML Framework each consisting of the components listed in Table~\ref{tab:components}. 
The first column lists different AI/ML components and the other two columns list the open-source tools utilized by the compared solutions: \textit{NAOMI} and \textit{O-RAN}.
The listed tools were presented and analyzed in Section~\ref{sec:tools}. In this section, we first elaborate on the experimental infrastructure and setup, summarized in Table~\ref{tab:scenarios_setups}, then we present the methodology used for evaluating NAOMI and comparing both solutions.
Further, we identify any limitations to the evaluation and results.

\begin{table}[htbp]
\caption{AI/ML Components for Evaluated Solutions}
\ra{1.2}
\centering
\fontsize{8pt}{10pt}\selectfont 
\begin{tabularx}{1\linewidth}{@{}>{\centering\arraybackslash}X|*{2}{>{\centering\arraybackslash}p{2.5cm}}@{}}
\toprule
\thead{\textit{\textbf{AI/ML Component}}} & \textit{\textbf{NAOMI}} & \textit{\textbf{O-RAN}} \\
\midrule\midrule
Data Preparation and Storage & Ray Data, MinIO & InfluxDB, Cassandra, \textit{Feature Store SDK*} \\
\midrule
Training & Ray (Train, Tune) & Kubeflow, \textit{Training Manager*} \\
\midrule
Model Management & MLflow & Leofs, \textit{Model Store SDK*} \\
\midrule
Model Deployment and Inference & Ray Serve & KServe \\
\midrule
Continuous Operation and Monitoring & Flyte, Prometheus, Grafana & Kubeflow, \textit{Dashboard Portal*} \\
\bottomrule
\multicolumn{3}{l}{\textit{*Tools built by O-RAN Software Community.}}
\end{tabularx}
\label{tab:components}
\end{table}

\begin{table*}[htbp]
\caption{AI/ML Model Training Scenarios}
\ra{1.2}
\centering
\fontsize{8pt}{10pt}\selectfont 
\begin{tabularx}{0.8\linewidth}{@{}>{\centering\arraybackslash}X|>{\centering\arraybackslash}p{3cm}|>{\centering\arraybackslash}p{2cm}|>{\centering\arraybackslash}p{4.5cm}@{}}
\toprule
\thead{\textit{\textbf{Scenario}}} & \textit{\textbf{Setup}} &  \textit{\textbf{Solution}} & \textit{\textbf{AI/ML Training}}\\
\midrule\midrule
Scenario 1* & Centralized & NAOMI & 1 Ray worker, 4 CPUs  \\
\midrule
Scenario 2 & Centralized & NAOMI & 1 Ray worker, 1 CPU  \\
\midrule
Scenario 3 & Distributed & NAOMI & 4 Ray workers, 1 CPU per worker  \\
\midrule
O-RAN SC Scenario* & Centralized & O-RAN & Up to 16 CPUs (\textit{Best Effort}) \\
\bottomrule
\multicolumn{4}{l}{*Solution's default scenario.}
\end{tabularx}
\label{tab:scenarios_setups}
\end{table*}

\subsection{Experimental Setup} 
\label{sec:exp-setup}

The experiments were performed on an infrastructure following two different setups.
The first setup (a centralized setup) is a single virtual machine with Ubuntu Linux following the specified minimal requirements of O-RAN AI/ML Framework~\cite{o-ran-sc-docs}.
The virtual machine consists of 16 x86\_64 architecture CPU cores and 32 GB of RAM.
VM processor is an Intel Xeon E5-2650 2.00 GHz CPU with one thread per core.

The second setup (a distributed edge setup) consists of the virtual machine listed in the preceding paragraph and three ARM architecture Raspberry Pi 5 nodes with 4 CPU cores and 8 GB of RAM each, which act as the edge devices connected to the same network over the Ethernet connection.
Raspberry Pi 5 nodes use 64-bit quad-core ARM Cortex-A76 processors with 2.4 GHz clock speeds.

Additional configurations are in place for ML model training.
Four scenarios and configurations for each solution are listed in Table~\ref{tab:scenarios_setups}.
The first scenario consists of a centralized setup with ML model training on one Ray worker with 4 CPU cores allocated.
Ray worker is a Kubernetes pod deployed on the cluster and is the main actor for ML model training.
The described configuration is the default configuration after NAOMI is deployed.
The second scenario is training on one Ray worker with one CPU core (a centralized setup).
The third scenario consists of distributed model training on 4 Ray workers with 1 CPU, where each worker is on a different node (a distributed edge setup).
In this scenario, the ML procedure needs to be modified to support distributed training over multiple Ray workers and physical nodes.
In all NAOMI scenarios, Ray workers are allocated 6 GB of RAM each.
The purpose of different scenarios for NAOMI is to evaluate the differences when distributing ML training over multiple CPU cores on a single node setup and a multiple node setup while comparing it to the baseline training on a single CPU core.

A default scenario configured by the O-RAN SC is training an ML model on up to 16 CPU cores and up to 32 GB of RAM (a centralized setup) as allocated by Kubernetes \textit{Best Effort} QoS (quality of service) class.
Pods with this QoS class are scheduled with a low priority and the allocated resources depend on the specified requests of other pods deployed on Kubernetes~\cite{KubernetesPodQoS}.

\subsection{Deployment Time}
\label{subsec:methods:deployment_time}

Deployment time of NAOMI and O-RAN solutions was measured on a centralized setup explained in Section~\ref{sec:exp-setup}.
Deployment time relates to service instantiating time, which, in the context of network services, is the time required for the provisioning and deployment of a network service~\cite{Khalili2019NetworkPlatforms}.

The set-up scripts were modified so that the deployment was fully automated for both solutions.
In addition, O-RAN AI/ML Framework scripts had to be modified with a different Kubernetes installation due to a dependency on a deprecated release.
The experiment was repeated five times for both solutions and each setup.

NAOMI supports deployment on a distributed setup, as described in Section~\ref{sec:exp-setup}. 
To test the deployment time of NAOMI on a distributed setup, we automated the deployment using an Ansible playbook.
We tested the deployment time from the time the Ansible playbook started executing to the time all the nodes and NAOMI deployments were in a ready state, providing insight into the overhead of a distributed system. To measure the deployment time over multiple runs, we define the average deployment time and its standard deviation as follows:
\begin{equation}
    T_{\text{dep}} = \bar{T} \pm \sigma_T, \quad \text{with}
\label{eq:deployment_time_1}
\end{equation}
\begin{equation}
    \bar{T} = \frac{1}{N}\sum_{i=1}^{N} T_i, \quad
    \sigma_T = \sqrt{\frac{1}{N-1}\sum_{i=1}^{N}\left(T_i - \bar{T}\right)^2},
\label{eq:deployment_time_2}
\end{equation}
where \(T_i\) is the measured deployment time for the \(i\)th run, \(N\) is the total number of runs, \(\bar{T}\) is the average deployment time, and \(\sigma_T\) is the standard deviation of the average deployment time. We computed the deployment time over $N = 10$ measurement runs and plotted the results and the standard deviation utilizing the above defined Eq. \ref{eq:deployment_time_1} and \ref{eq:deployment_time_2}.

\subsection{Resource Usage}
\label{subsec:methods:resource_usage}

Kubernetes CPU and RAM requests and limits were collected after the deployment of both solutions.
Configurations for both solutions were kept at default settings.
Both systems were deployed on a centralized setup with their default scenario as presented in Section~\ref{sec:exp-setup}.
Additionally, the CPU and RAM utilization of the virtual machine were collected in an idle state.
For collecting metrics, the \textit{kubectl} CLI tool was utilized for Kubernetes requests and limits, and the \textit{free -m} tool for RAM usage on the virtual machine.

For scheduling and placing deployments on a Kubernetes cluster, requests and limits need to be specified.
They specify how much of each resource a container needs and a Kubernetes scheduler (\textit{kube-scheduler)} uses this to decide where to place the deployment.
If limits are specified as higher than requests, a container is allowed to use more than its requests but will not get scheduled if there are fewer resources available than specified in the requests~\cite{KubernetesRes}.

During the example AI/ML workflow execution, resource utilization was measured at 1 s intervals and CPU usage and RAM usage graphs for both solutions were plotted with timestamps indicating the start of each of the workflow stages. We calculated the average utilization for CPU and RAM during each workflow stage from the measured resource time series. The average utilization for each stage was computed using the following equation across all measurement points for each of the workflow stages:

\begin{equation}
\bar{R}_j = \frac{1}{M_j} \sum_{k=1}^{M_j} R_{j,k},
\label{eq:resource_usage}
\end{equation}

where \(\bar{R}_j\) represents the average resource usage (CPU or RAM) during the \(j\)th workflow stage, \(M_j\) is the total number of measurements in stage \(j\), and \(R_{j,k}\) is the resource usage measured at the \(k\)th interval.

\subsection{Example AI/ML Workflow Execution Length} 
\label{subsec:methods:workflow_exec}

O-RAN SC provides an example AI/ML workflow for training a quality of experience (QoE) prediction model.
The QoE workflow includes feature creation, data preparation, and training of a deep learning model for a QoE prediction service.
They also provide a dataset example of time series measurements of network metrics such as the amount of uploaded and downloaded \textit{pdcp} bytes\, throughput, and network cell identity~\cite{o-ran-sc-docs}.
To evaluate the solutions under different loads, we scaled the dataset 10 and 100 times.

The QoE workflow was modified to work with NAOMI for all scenarios and setups.
We measured the entire workflow execution time, as well as data extraction, model training, model deployment, and overhead tasks.
For fair evaluation we did not utilize NAOMI's Ray Data tool, but the same data preparation code with the Pandas framework for data preparation as the QoE workflow provided by O-RAN SC.
NAOMI automates the model deployment component however, in the O-RAN solution this is a manual process.
Considering that the model deployment is a crucial component of the AI/ML workflow, as explained in Section~\ref{subsec:components:inference}, we included it in the results.

For QoE model training, we used the default hyperparameters provided by O-RAN SC, which are \textit{batch size 10} and \textit{epoch 1}.
Model training was done on scenario 1 for NAOMI and on the O-RAN SC scenario as described in Section~\ref{sec:exp-setup}.

The entire QoE workflow execution time was measured from job initialization to finished response for both solutions and three different dataset sizes.
We developed an interface for sending job and status requests for the O-RAN solution because their solution supports only the user interface (UI) through a web browser for starting the jobs.
The execution time of each stage of the workflow was also collected, while all the times were collected as reported by the user interfaces of both solutions.
These times also include overhead tasks like container creation and initialization.
All the experiments were repeated five times. 
The average times per stage and the variations of the results are plotted in the figure.

We utilized Eq. \ref{eq:deployment_time_1} and Eq. \ref{eq:deployment_time_2} formalized in Section~\ref{subsec:methods:deployment_time} to compute the average execution time for each workflow stage and the overall workflow, as well as their corresponding standard deviations. These metrics were calculated across all 5 runs and the results are presented with their respective standard deviation bars in the graphs.

\subsection{Distributed Training} \label{subsec:methods:dist:train}

As discussed in Section~\ref{sec:solution}, unlike the O-RAN solution, NAOMI supports distributed training. 
When distributing the model training, the part of the model training workflow that can be parallelized can be sped up by a factor equal to the number of CPU cores, while the entire training procedure speedup is limited by the non-parallel portion~\cite{Hill2008AmdahlsEra}.
However, distributed systems include some overhead for managing distributed tasks and communication between nodes~\cite{Jogalekar2000EvaluatingSystems}.
In this evaluation, we focus on collecting results of distributed system overhead for workflow deployment and usage of resources per node during the training.

To evaluate the performance of NAOMI in a distributed set-up and gain insight into the overhead, we modified the QoE workflow, introduced in Section~\ref{subsec:methods:workflow_exec}, to support distributed training with the Ray Framework component, which is model training component in NAOMI, as explained in Section~\ref{sec:solution} and depicted in Fig.~\ref{fig:nancy-arch}.

Batch size affects the performance of the distributed training~\cite{You2019Large-batchBeyond}, so training time was measured for different batch sizes and compared between different scenarios listed in Section~\ref{sec:exp-setup}.
For a comparison, we included training time on O-RAN AI/ML Framework with the default scenario.
Resource usage metrics and training time were measured in the same manner as described in Section~\ref{subsec:methods:resource_usage}. We utilized Eq.~\ref{eq:deployment_time_1} formalized in Section~\ref{subsec:methods:deployment_time} to compute the average training time $\bar{T}_{s}(B)$ as a function of batch size and scenario in Table~\ref{tab:scenarios_setups}.

\subsection{Inference Latency}
\label{subsec:methods:inference}

After the model is trained it is deployed as an API endpoint and inference can be performed on the model, as discussed in Section~\ref{subsec:components:inference}.
In the use case provided by the O-RAN, presented in Section~\ref{subsec:methods:workflow_exec}, quality of experience (QoE) prediction is performed.
To evaluate the inference performance of the solutions, the inference service was tested using 1 to 1024 concurrent requests, doubling the number of concurrent requests in each step and the average end-to-end delay was measured.

Both tools for the model inference component (Ray Serve for NAOMI and KServe for O-RAN) utilize a horizontal autoscaler for increasing the number of model replicas as the load increases.
Configurations for both autoscalers were kept at default for a fair comparison of the inference service.

As NAOMI supports deployment on a distributed setup, described in Section~\ref{sec:exp-setup}, the inference was evaluated for NAOMI when deployed on such setup.
We evaluate the ability of load-balancing on multiple nodes and any improvements in mean latency as inference is performed distributively across more CPU cores.
On a distributed setup there are 16 cores available for model inference compared to 4 cores on a centralized setup.

To formalize the impact of parallel processing on inference latency, we decompose the overall mean latency \(\bar{L}\) into two components: a serial component that includes network communication and other system overhead, and a parallelizable component corresponding to the distributed computation. Referring to Amdahl's findings~\cite{Amdahl1967ValidityCapabilities, Hill2008AmdahlsEra} this is expressed as:
\begin{equation}
\bar{L} = L_{\text{serial}} + \frac{L_{\text{parallel}}}{n},
\label{eq:latency}
\end{equation}
where \(L_{\text{serial}}\) represents the portion of work that can not be parallelized such as network communication, scheduling, and other overhead work, \(L_{\text{parallel}}\) denotes the latency of tasks that can be executed in parallel, and \(n\) is the number of CPU cores available for inference.

The theoretical speedup \(S\) when increasing the available cores from 4 to 16 is given by the ratio:
\begin{equation}
S = \frac{L_{\text{serial}} + \frac{L_{\text{parallel}}}{4}}{L_{\text{serial}} + \frac{L_{\text{parallel}}}{16}}.
\label{eq:speedup}
\end{equation}
In the ideal case where \(L_{\text{serial}}\) is negligible compared to \(L_{\text{parallel}}\), Eq.~\ref{eq:speedup} simplifies to:
\begin{equation}
S \approx \frac{\frac{L_{\text{parallel}}}{4}}{\frac{L_{\text{parallel}}}{16}} = \frac{16}{4} = 4,
\label{eq:speedup_actual}
\end{equation}
resulting in a theoretical speedup of 4 times.

\subsection{Heterogeneous Scale Study}
\label{subsec:methods:scale_study}

As shown in Section~\ref{sec:solution} and discussed in Sections~\ref{subsec:methods:dist:train} and ~\ref{subsec:methods:inference}, NAOMI supports distributed training and scalable inference across multiple heterogeneous nodes. To evaluate NAOMI's scalability under  real-world network conditions, we conducted a set of scale study experiments utilizing the devices presented in Section~\ref{sec:exp-setup}. 
Our test environment consisted of an x86\_64 virtual machine, three ARM-based Raspberry Pi nodes as presented in Section~\ref{sec:exp-setup}, and an additional Intel virtual machine (12 CPU cores, 32 GB RAM) was added to the distributed setup. Nodes were connected through Ethernet cables, creating a representative multi-device distributed and heterogeneous network computing environment.

The scale study increased the number of CPU cores and Ray workers to scale up the computational power of the Ray cluster. 
We measured QoE model inference performance on parallel requests as discussed in Section~\ref{subsec:methods:inference}. Model inference is parallelizable and can potentially benefit from a distributed setup as discussed in the preceding sections.

We calculated ideal speedup using Eq. \ref{eq:latency}, \ref{eq:speedup}, and \ref{eq:speedup_actual} and measured the actual speedup achieved by increasing the number of CPU cores for model inference. Model replicas were scaled along with CPU cores and Ray workers, each model replica utilizing 1 CPU core. Replicas were distributed along the physical nodes by Ray Serve library and Kubernetes. The measured results and ideal speedup for heterogeneous scale study were plotted on the graph.

\subsection{Workflows Beyond O-RAN}
\label{subsec:methods:general_models}

NAOMI supports a wide variety of AI/ML workflows and models beyond network use cases, as discussed in Section~\ref{subsec:general_applica}. 
To elaborate on those claims, we adapted three additional workflows for NAOMI. 
Two models from MLPerf Tiny Benchmark for benchmarking inference on low-powered devices~\cite{banbury2021mlperf}.
Specifically, Image Classification ResNetv1~\cite{He2016DeepRecognition} model trained on CIFAR-10~\cite{Krizhevsky2009CIFAR-10Datasets} dataset and Visual Wake Words MobilenetV1~\cite{howard2017mobilenetsefficientconvolutionalneural} model trained on MSCOCO 2014~\cite{Lin2014MicrosoftContext} dataset. However, the MSCOCO dataset had to be reduced to 10\% of its original size due to limitations in our testbed infrastructure.
Further, we adapted a common introductory ML model for MNIST handwritten digit classification~\cite{Kadam2020CNNDataset} to be trained and deployed on NAOMI.

The datasets are of different sizes and differ in the number of samples, which affects data extraction and training time. Image Classification CIFAR-10 is $\approx$740 MB with 60000 samples. The Visual Wake Words MSCOCO dataset due to its size was reduced to 1.1 GB with 10000 images. The QoE example time series dataset from O-RAN has $\approx$0.09 MB with 1029 samples, and lastly, MNIST has $\approx$25 MB with 42000 images.

These three models were evaluated on Scenario 1 from Table~\ref{tab:scenarios_setups}. The Pandas framework was used for data extraction except for MNIST, where Ray Data was utilized. 
Execution length was measured in the same manner as discussed in Section~\ref{subsec:methods:workflow_exec}, and inference latency was evaluated using the methodology presented in Section~\ref{subsec:methods:inference}.
These results were compared to the QoE workflow example evaluation discussed in the preceding sections.

\section{Evaluation Results} 
\label{sec:results}

In this section, we present and discuss evaluation results for the proposed \textit{NAOMI}, a solution introduced in Section~\ref{sec:solution} according to the evaluation methodology detailed in Section~\ref{sec:methodology}. In the following paragraphs we first measure the overall system deployment time and required CPU and memory resources, followed by per step duration of the ML workflow execution. Next, we focus on the workflow execution in a distributed setup uniquely enabled by NAOMI and finally, we focus on inference performance.

\subsection{System Deployment}
\begin{figure}[htbp]
\centerline{\includegraphics[width=1\linewidth]{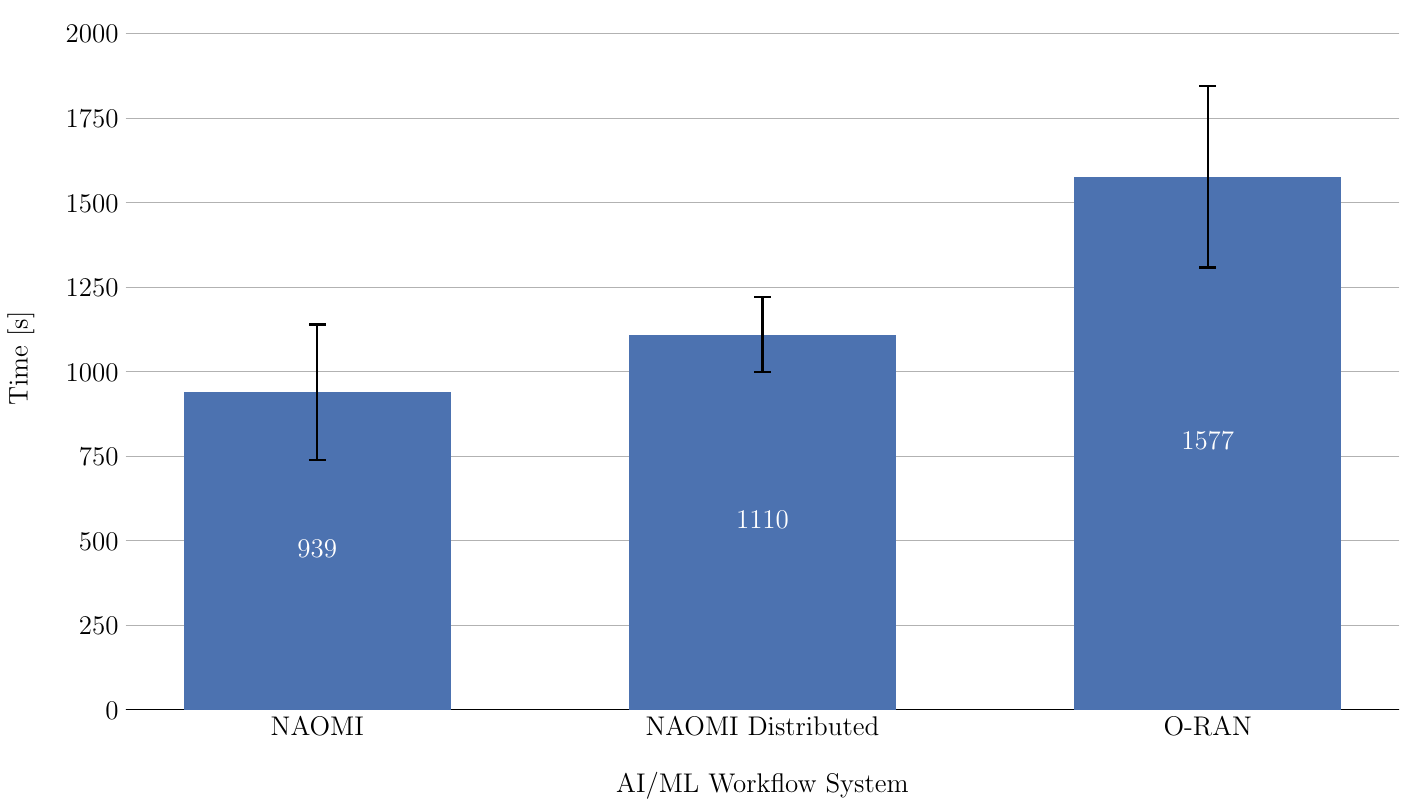}}
\caption{Time required for each AI/ML workflow system to be deployed.}
\label{fig:dep_time}
\end{figure}

Fig.~\ref{fig:dep_time} presents the deployment time of NAOMI and O-RAN solutions on a centralized setup along with the deployment time of NAOMI on a distributed setup, as explained in Section~\ref{subsec:methods:deployment_time}.
The deployment time is plotted on the \(y\) axis, the average time is represented by the height of the box plot while the standard deviation is represented by the corresponding interval line.

From the figure, we can see that NAOMI on average takes $939 \pm 200s$ and $1100 \pm 110s$ to deploy in centralized and distributed setups respectively, while O-RAN requires $1577 \pm 270s$. In other words, NAOMI requires $40\%$ less time ($\Delta 638 s$) to deploy compared to O-RAN's and $30\%$ less ($\Delta 467 s$) if we deploy NAOMI on a distributed setup.
The overhead of deploying NAOMI on a distributed setup compared to a centralized setup is $171s$ on average. There are significant variations in the deployment time for all evaluated deployments.
For O-RAN the standard deviation is $\approx$$270 s$ and for NAOMI $\approx$$200 s$ on a centralized setup.
These variations are expected due to the internet bandwidth, as pulling container images from repositories and installing software is a major part of deployment scripts for both solutions.

Although deploying NAOMI on a distributed setup (4 nodes compared to 1 node) adds overhead, it is not substantial and validates NAOMI's scalability on a distributed network edge.
A rapid deployment time is important for fast instantiation of services on network slices.
Deploying an AI/ML workflow system as a service and upgrading existing deployments quicker relates to the service instantiation time KPI on networks~\cite{Khalili2019NetworkPlatforms}, as discussed in Section~\ref{subsec:methods:deployment_time}.

\begin{figure}[htbp]
\centerline{\includegraphics[width=1\linewidth]{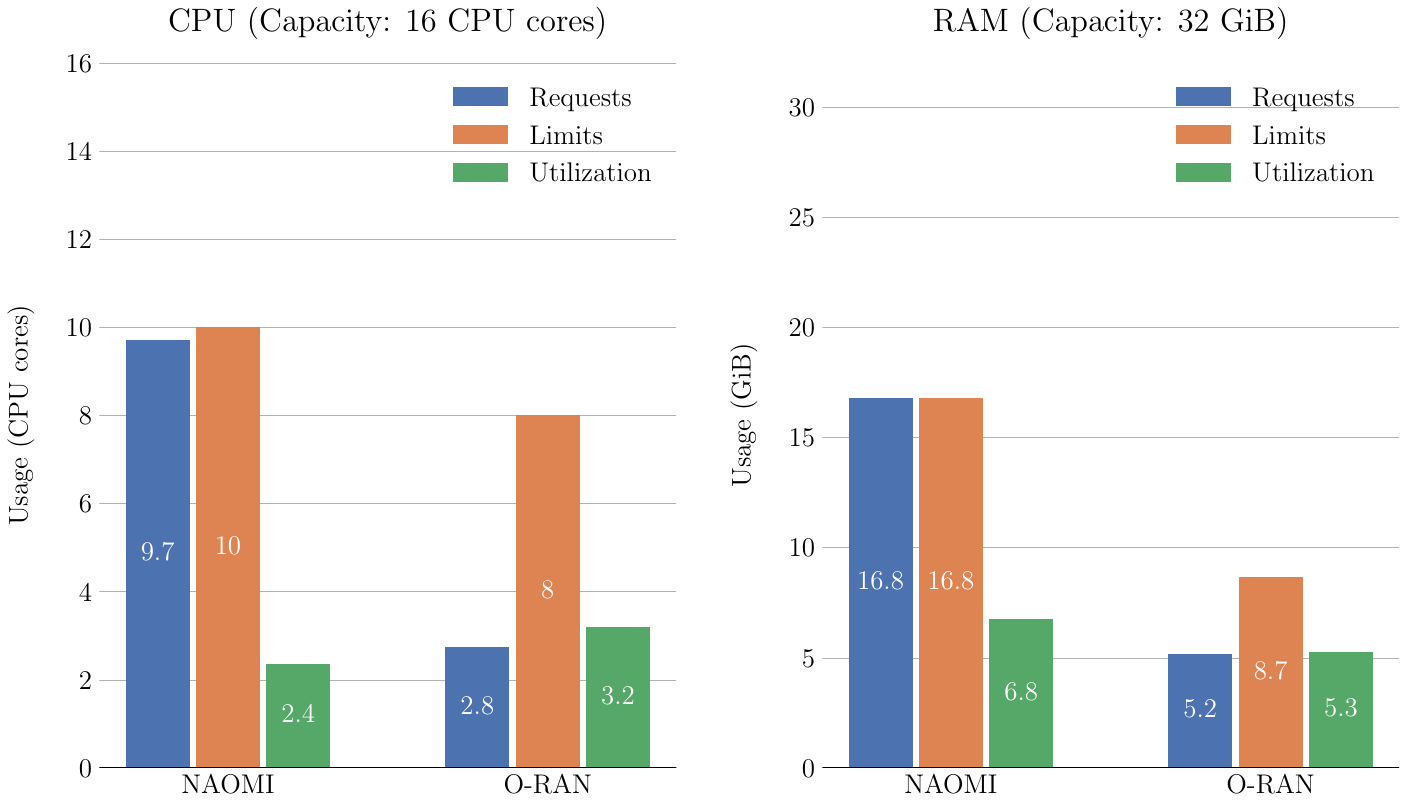}}
\caption{Kubernetes requests, limits, and usage of resources of AI/ML workflow solutions in an idle state.}
\label{fig:res_usage}
\end{figure}

Requests and limits of resources and utilization of resources on the virtual machine were collected for both solutions, as explained in Section~\ref{subsec:methods:resource_usage} and are presented in Fig.~\ref{fig:res_usage}.
Figures for CPU and RAM present metrics on the \(y\) axis, where the maximum value of the axis is the capacity of the virtual machine.
The utilization column can exceed the requests column due to other processes running on the machine.

From the figure, we can see that NAOMI limits are set to 10 CPU cores and 16.8 GiB of RAM, with CPU requests lower at 9.7 cores. O-RAN requests are 2.8 cores and 5.2 GiB for CPU and RAM respectively, with limits set to 8 CPU cores and 8.7 GiB of RAM.
The utilization of CPU for NAOMI is 2.4 cores, while for O-RAN it is 3.2 cores, and for NAOMI utilizes 6.8 GiB of RAM, compared to O-RAN, which uses 5.3 GiB.
O-RAN CPU limits are set to 20\% lower than NAOMI, however the virtual machine utilization of CPU is 25\% lower for NAOMI.
The difference in RAM requests is larger, with O-RAN requesting $\approx$48\% less RAM and utilizing $\approx$22\% less in an idle state compared to NAOMI.

These results indicate that despite NAOMI allocating more resources, it utilizes less CPU in an idle state.
Higher requests, limits, and RAM utilization are due to deploying more open-source components, such as Grafana and Prometheus monitoring software, which O-RAN AI/ML Framework does not utilize.

\subsection{QoE AI/ML Workflow Execution}
\label{subsec:results:workflow_exec}

\begin{figure}[htbp]
\centerline{\includegraphics[width=1\linewidth]{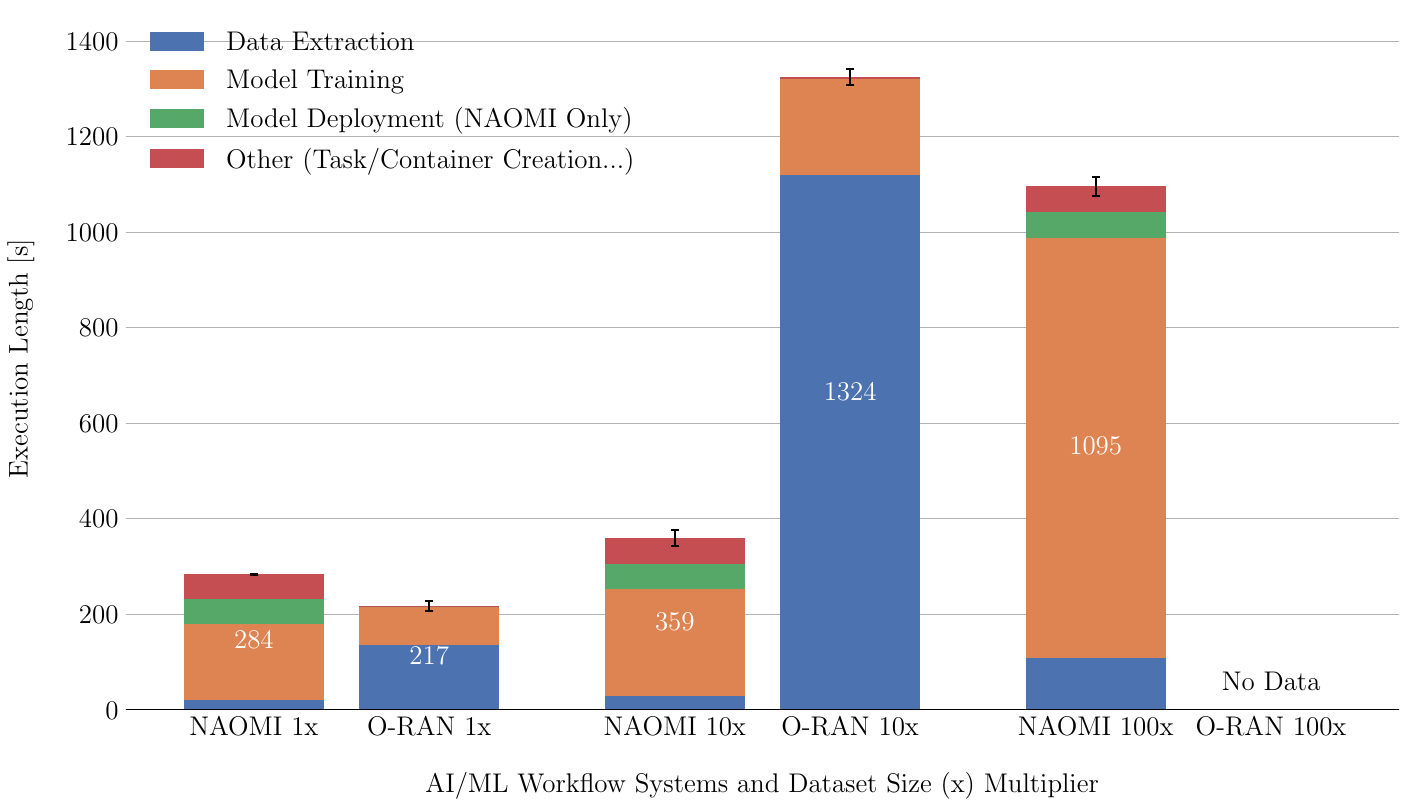}}
\caption{Execution time for each stage of the AI/ML workflow.}
\label{fig:exec_time}
\end{figure}

Presented in Fig.~\ref{fig:exec_time} are the execution times of the QoE prediction workflow on both solutions for a reference dataset ($1x$), a dataset with 10 times the size of the reference dataset ($10x$) and a dataset 100 times the size of the reference one ($100x$).
The results correspond to experiments according to Scenario 1 for NAOMI and the default scenario for O-RAN, as explained in Section~\ref{subsec:methods:workflow_exec} and Table~\ref{tab:scenarios_setups}.
Different dataset sizes and solutions are plotted on the \(x\) axis and execution time in seconds on the \(y\) axis for each of the experiments.
Each column consists of sections that present the execution time of workflow stages and any overhead tasks, such as container creation, labeled as \textit{Other}.

For the $1x$ dataset, the O-RAN solution outperforms NAOMI by $\approx$24\%, however, the automated model deployment stage is not executed in O-RAN.
Excluding the model deployment component the difference is marginal at $\approx$$14 s$.
Model training is around $50 \%$ faster on O-RAN, while the data extraction phase takes approximately $84\%$ less time on NAOMI for a 1x dataset size.
From these results, we can see that NAOMI performs data extraction considerably better, while it brings more overhead with model training due to sending jobs to the Ray cluster.
NAOMI has more overhead tasks like container and task creation, however, the discrepancy could be due to different approaches to measuring each stage execution time reported by the user interfaces, as explained in Section~\ref{subsec:methods:workflow_exec}.

It can also be noticed in the figure that NAOMI performs significantly better when the dataset size increases, specifically in the data extraction component of the AI/ML workflow, where for the $10x$  dataset, NAOMI executes data extraction $\approx$97\% faster.
While NAOMI data extraction is considerably faster, MinIO, the selected tool for the data preparation component, does not support structured data queries like InfluxDB, as discussed in Section~\ref{subsec:components:data_prep}.

The relative difference in model training is lower for the $10x$ dataset, where O-RAN performs around 10\% better than NAOMI.
The entire workflow execution time for the $10x$ dataset is $\approx$73\% faster on NAOMI.
Model training is executed faster on the O-RAN solution for each of the dataset sizes due to the overhead of sending ML training jobs to the Ray Cluster on NAOMI and the \textit{Best Effort} quality of service setting for O-RAN Kubeflow ML training pods, as explained in Section~\ref{sec:exp-setup}. 
The O-RAN solution can use up to 16 CPU cores for model training, whereas NAOMI can use 4 cores, as shown in Table~\ref{tab:scenarios_setups}.

Further, the O-RAN solution did not complete the execution of the workflow with the $100x$ dataset, presumably due to timeouts in data extraction stages, which were observed in the logs. 
NAOMI completed the execution in $1095 \pm 20s$ with $100x$ dataset size, which is approximately 3 times longer than the $10x$ dataset.
The difference in execution times for NAOMI between the $1x$ and $10x$ datasets is 75 seconds, and between the $10x$ and $100x$ datasets, it is 736 seconds. 
This indicates that the portion of the workflow execution that takes longer as the dataset size grows increases roughly linearly with the size of the dataset.

The execution time of other, overhead tasks for NAOMI took $\approx$$53 s$ for all three dataset sizes, while for O-RAN it took $\approx$$3 s$ for the $1x$ and $10x$ dataset, as can be seen in Fig.~\ref{fig:exec_time}.
Similarly, the deployment time on NAOMI took $\approx$$52 s$ for all three dataset sizes, while O-RAN does not support automated model deployment as discussed in Section~\ref{subsec:methods:workflow_exec}.
As we can see, the execution time of other, overhead tasks, and the model deployment stage did not increase as a function of the dataset multiplier, which indicates that dataset size has little to no effect on these AI/ML workflow stages.

\begin{figure}[!htb]
    \centering
    \begin{subfigure}[t]{\linewidth}
        \centering
        \includegraphics[width=1\linewidth]{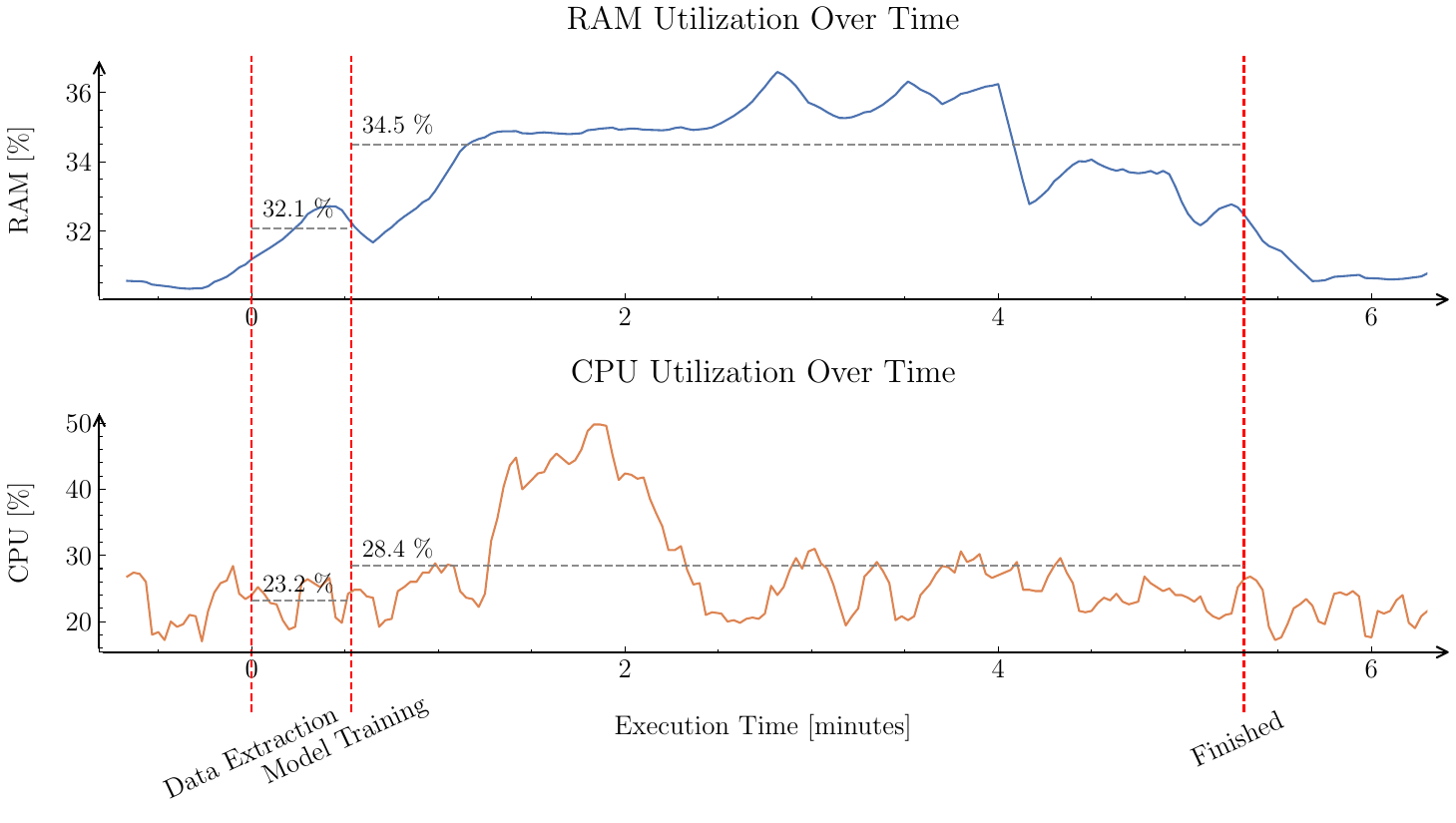}
        \caption{NAOMI}
        \label{subfig:exec_usage_nancy}
        \vspace{10pt}
    \end{subfigure}
        \begin{subfigure}[t]{\linewidth}
        \centering
        \includegraphics[width=1\linewidth]{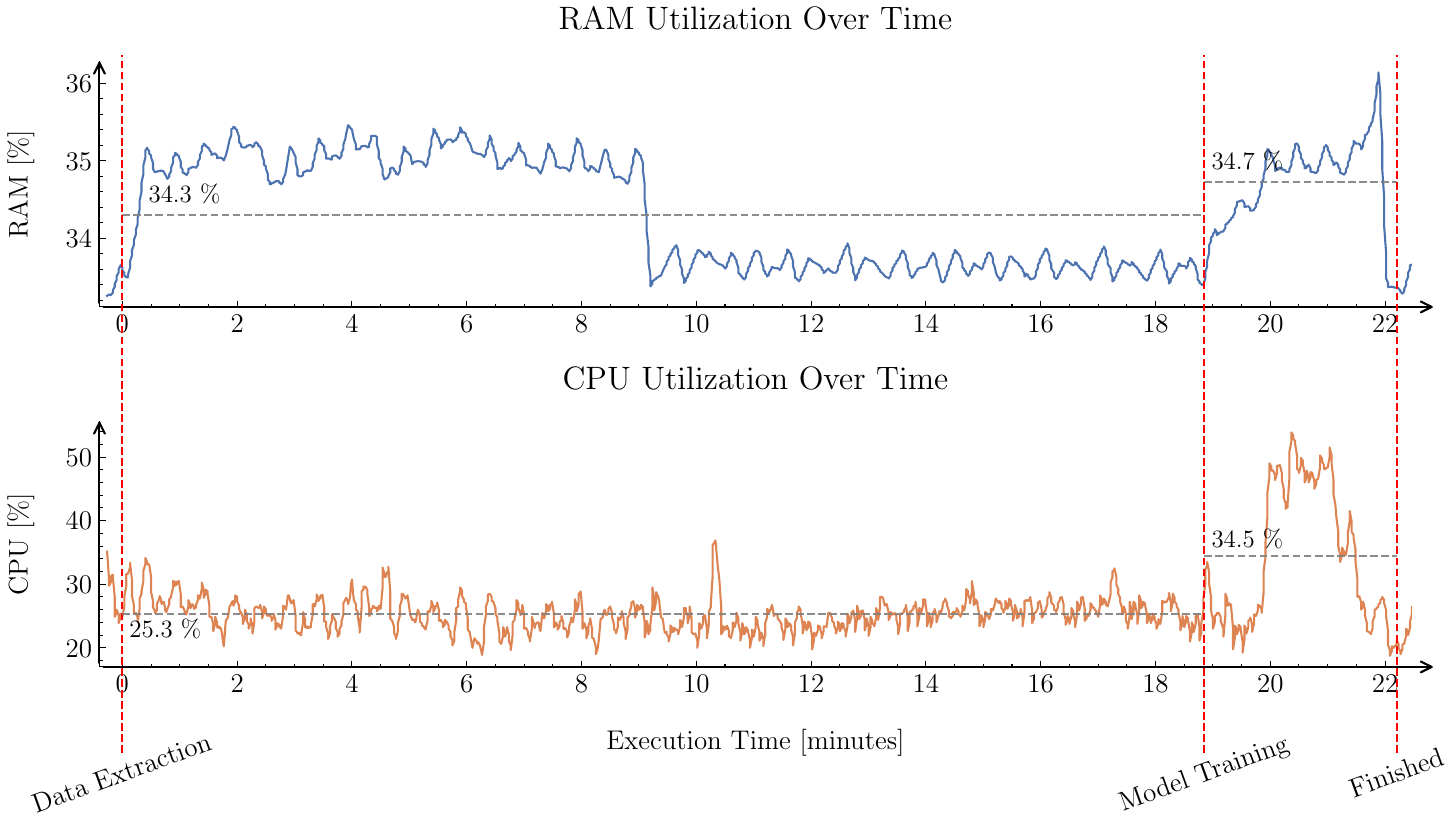}
        \caption{O-RAN AI/ML Framework}
        \label{subfig:exec_usage_oran}
        \vspace{20pt}
        
    \end{subfigure}
    \caption{Resource utilization during workflow execution.}
    \label{fig:exec_usage_combined}
\end{figure}

Resource utilization during the workflow execution with the $10x$ dataset is plotted in Fig.~\ref{fig:exec_usage_combined}.
Graphs are plotted for both resource types with NAOMI illustrated in Fig.~\ref{subfig:exec_usage_nancy} and O-RAN in Fig.~\ref{subfig:exec_usage_oran}.
Timestamps on the \textit{x} axis indicate the start and finish of data extraction and model training.
The average resource utilization is marked for both stages. 

NAOMI on average utilizes $32.1\%$ of RAM and $23.2\%$ of the entire CPU capacity during data extraction. 
During model training, it utilizes $34.5\%$ of RAM and $28.4\%$ of CPU.
O-RAN, on the other hand, utilizes $34.3\%$ of RAM and $25.3\%$ of CPU during data extraction, while it utilizes $34.7\%$ of RAM and $34.5\%$ of CPU during model training.

From Fig.~\ref{fig:exec_usage_combined} we can see that the O-RAN solution utilizes 2 percentage points more RAM and CPU during the data extraction phase.
Both workflows are not limited by CPU during data extraction, as the difference in CPU utilization during that stage compared to the idle state is insignificant.
Further, there is a clear distinction between RAM utilization in the two parts of the data extraction component for O-RAN from $0$ to $\approx$$9$ minutes and from $\approx$$9$ to $\approx$$19$ minutes, where 0 is marked as the start of the data extraction, as we can see from Fig.~\ref{subfig:exec_usage_oran}.
As mentioned in Section~\ref{sec:workflow_architecture}, the O-RAN solution uses InfluxDB as a data lake and Cassandra as a feature store.
From the system logs, we confirmed that operations from $0$ to $\approx$$9$ minutes are related to InfluxDB data extraction operations, while from $\approx$$9$ to $\approx$$19$ minutes, they relate to storing and extracting features from Cassandra.
This also indicates that Cassandra doesn't utilize a measurable amount of RAM in this setup, relying only on disk I/O operations, possibly slowing down the data extraction phase.
In NAOMI, MinIO brings a simple solution to extracting data and storing features, which for this use case provides better performance while utilizing a similar amount of resources as the O-RAN data extraction phase.

For model training, marked with a timestamp in Fig.~\ref{fig:exec_usage_combined}, the average CPU utilization is 6 percentage points higher on O-RAN and the difference in RAM utilization is within the margin of error.
For this scenario, model training takes a similar length of time while using less CPU on NAOMI.
This is due to NAOMI limiting model training to 4 CPU cores in this scenario and O-RAN utilizing up to 16 cores, as explained in Section~\ref{sec:exp-setup}.

Overall, the differences in resource utilization are comparable, while NAOMI executes the entire QoE workflow faster.
During the entire workflow execution, resource utilization does not exceed $50\%$, indicating that there are available resources on a physical machine for scheduling other tasks during the workflow execution for both solutions.

\subsection{Distributed Training}

\begin{figure}[htbp]
\centerline{\includegraphics[width=1\linewidth]{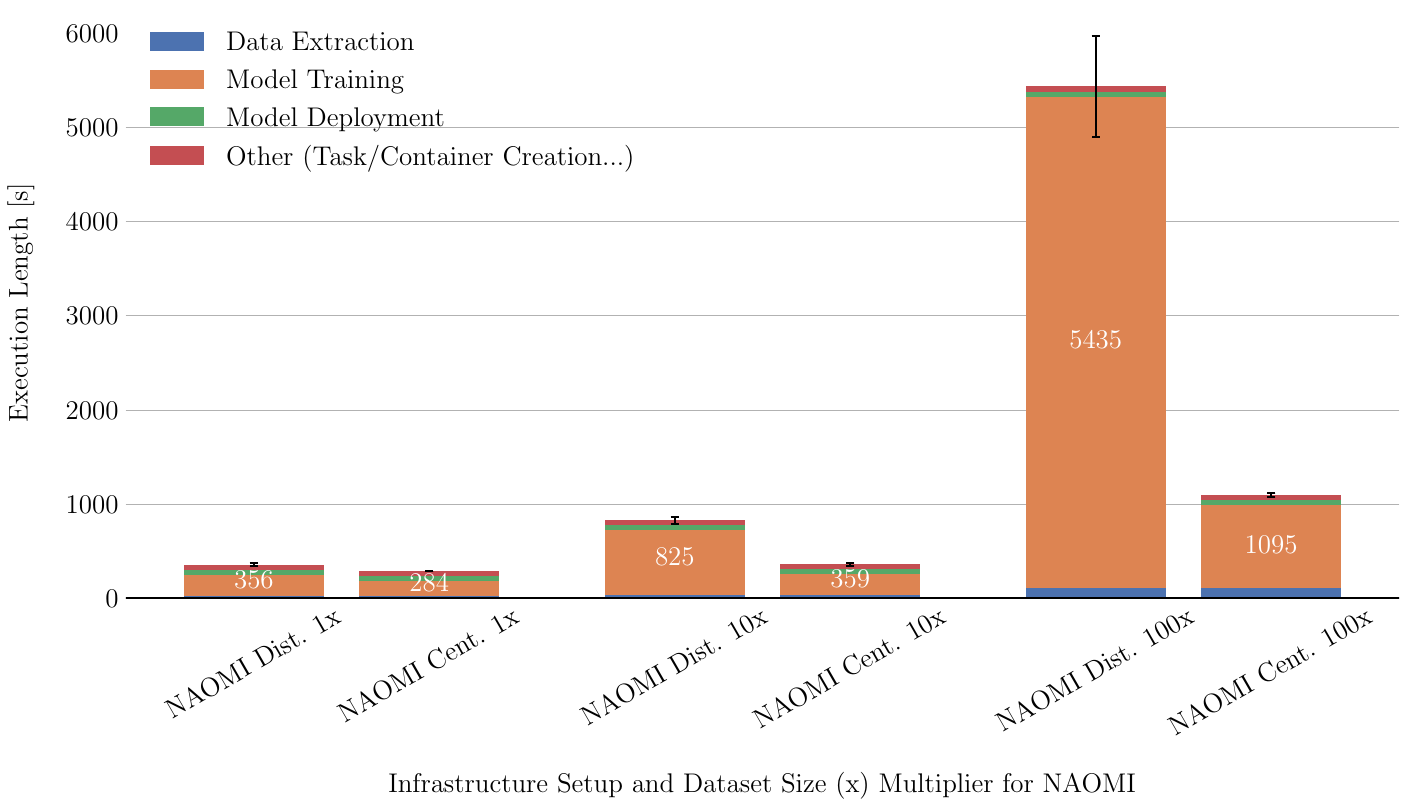}}
\caption{Execution time for each stage of the AI/ML workflow comparing distributed and centralized setups.}
\label{fig:dist_exec_time}
\end{figure}

Deploying NAOMI on a distributed setup with Scenario 3 provides a configuration with a similar amount of allocated resources for model training to Scenario 1, while the CPU cores are distributed over 4 physical nodes, as explained in Section~\ref{sec:exp-setup} and summarized in Table~\ref{tab:scenarios_setups}.
Fig.~\ref{fig:dist_exec_time} depicts the execution times of QoE prediction workflow execution comparing NAOMI deployed on a distributed and centralized setup.

Similarly to Fig.~\ref{fig:exec_time} in Section~\ref{subsec:results:workflow_exec}, Fig~\ref{fig:dist_exec_time} present execution times for a reference dataset (1x), a dataset with 10 times the size of the reference dataset (10x) and with a dataset 100 times the size of the reference one (100x). 
The results correspond to experiments according to Scenario 1 for NAOMI and Scenario 3 for NAOMI distributed as listed
in Section~\ref{sec:exp-setup} and Table~\ref{tab:scenarios_setups}. 
Each column consists of sections that present the execution time of workflow stages, such as data extraction, model training, and model deployment.

From Fig.~\ref{fig:dist_exec_time}, we can see that NAOMI on a centralized setup executed the QoE workflow in $284 \pm 1s$, $359 \pm 17s$, and $1095 \pm 20s$ for $1x$, $10x$, and $100x$ dataset respectively, while NAOMI on a distributed setup executed the QoE workflow in $356 \pm 17s$, $825 \pm 37s$, and $5435 \pm 539s$ for described dataset sizes.
For model training the distributed setup together with distributed model training performs significantly worse.
The centralized setup performed $31\%$ faster for $1x$ dataset size, $68\%$ for $10x$ and $83\%$ for $100x$.
This shows a significant overhead of distributed model training for this set of training parameters. 
The distributed setup does not affect data extraction, model deployment, and \textit{other} (overhead) tasks.
In all cases, both setups performed these tasks with a difference of up to 3 seconds, which is within the margin of error.

Distributed model training performing worse than normal training occurs due to the additional communication between nodes and workers in the distributed setup, which increases processing time, particularly with larger datasets and low batch sizes. Synchronizing model updates across the distributed workers introduces delays. As depicted in Fig.~\ref{fig:dist_exec_time}, when the dataset size increases, we can observe a bigger performance gap between the centralized and distributed setups.

\begin{figure}[htbp]
\centerline{\includegraphics[width=1\linewidth]{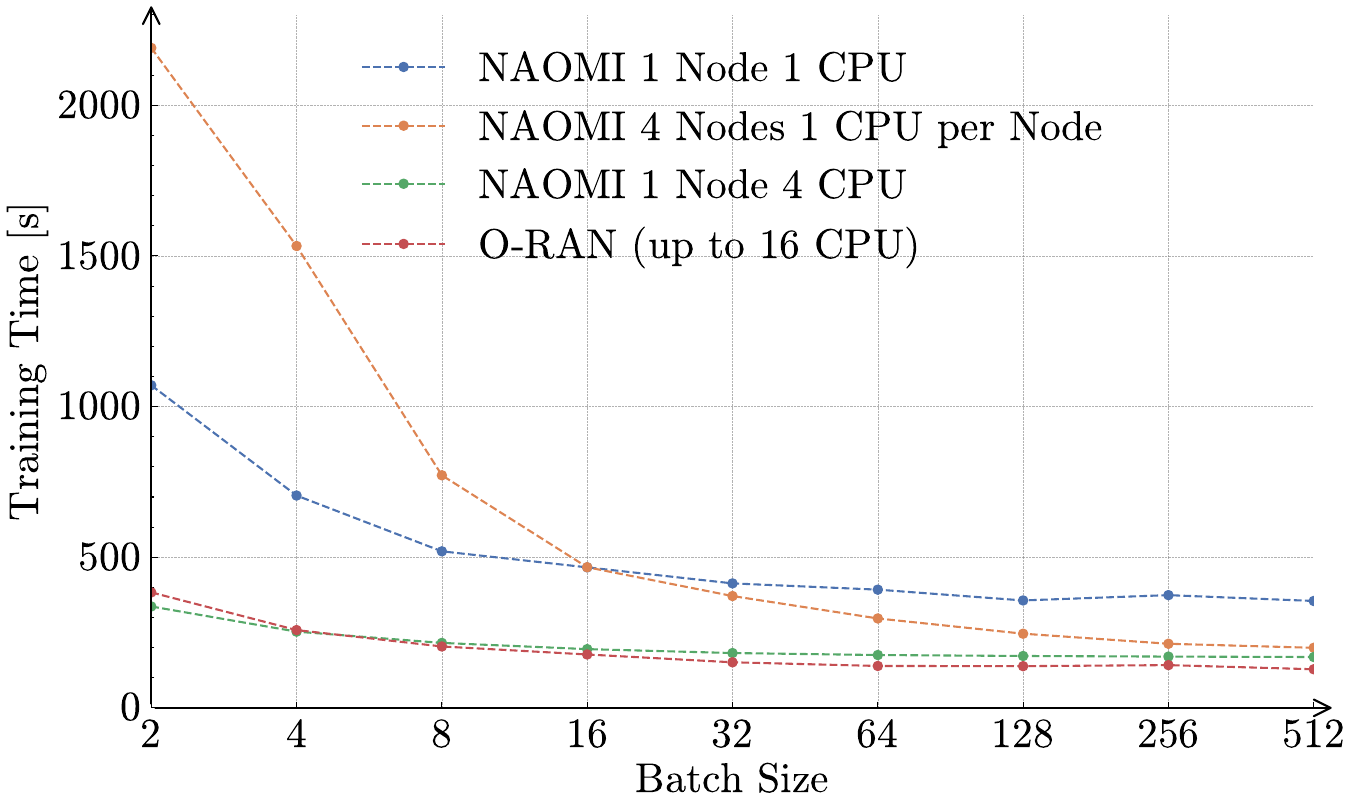}}
\caption{Training time per batch size for different scenarios summarized in Table~\ref{tab:scenarios_setups}.}
\label{fig:train_batch}
\end{figure}

Given that batch size impacts distributed model training speed, as discussed in Section~\ref{subsec:methods:dist:train}, a further evaluation of the distributed and centralized training is performed as a function of increasing batch size.
Fig.~\ref{fig:train_batch} shows model training time ($y$ axis) for different batch sizes ($x$ axis), where $x$ axis is in logarithmic scale.
Training time is presented for 3 different scenarios on NAOMI as explained in Section~\ref{sec:exp-setup} and compared to the O-RAN default scenario.

\begin{figure}[!htb]
    \centering
    \begin{subfigure}[htbp]{\linewidth}
        \centering
        \includegraphics[width=1\linewidth]{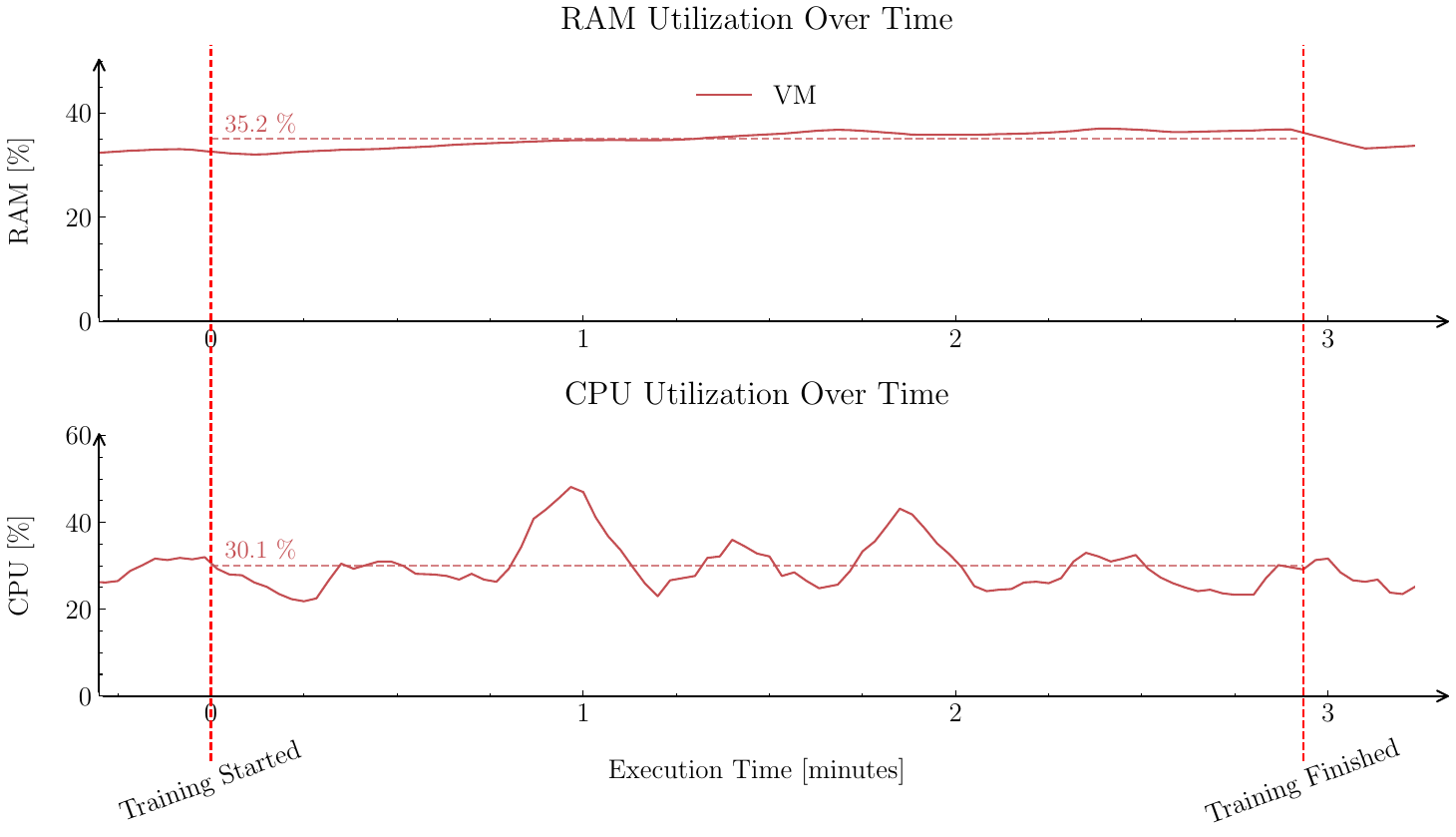}
        \caption{Single node training (Scenario 1).}
        \label{subfig:single_node_training}
        \vspace{20pt}
        
    \end{subfigure}
    \begin{subfigure}[htbp]{\linewidth}
        \centering
        \includegraphics[width=1\linewidth]{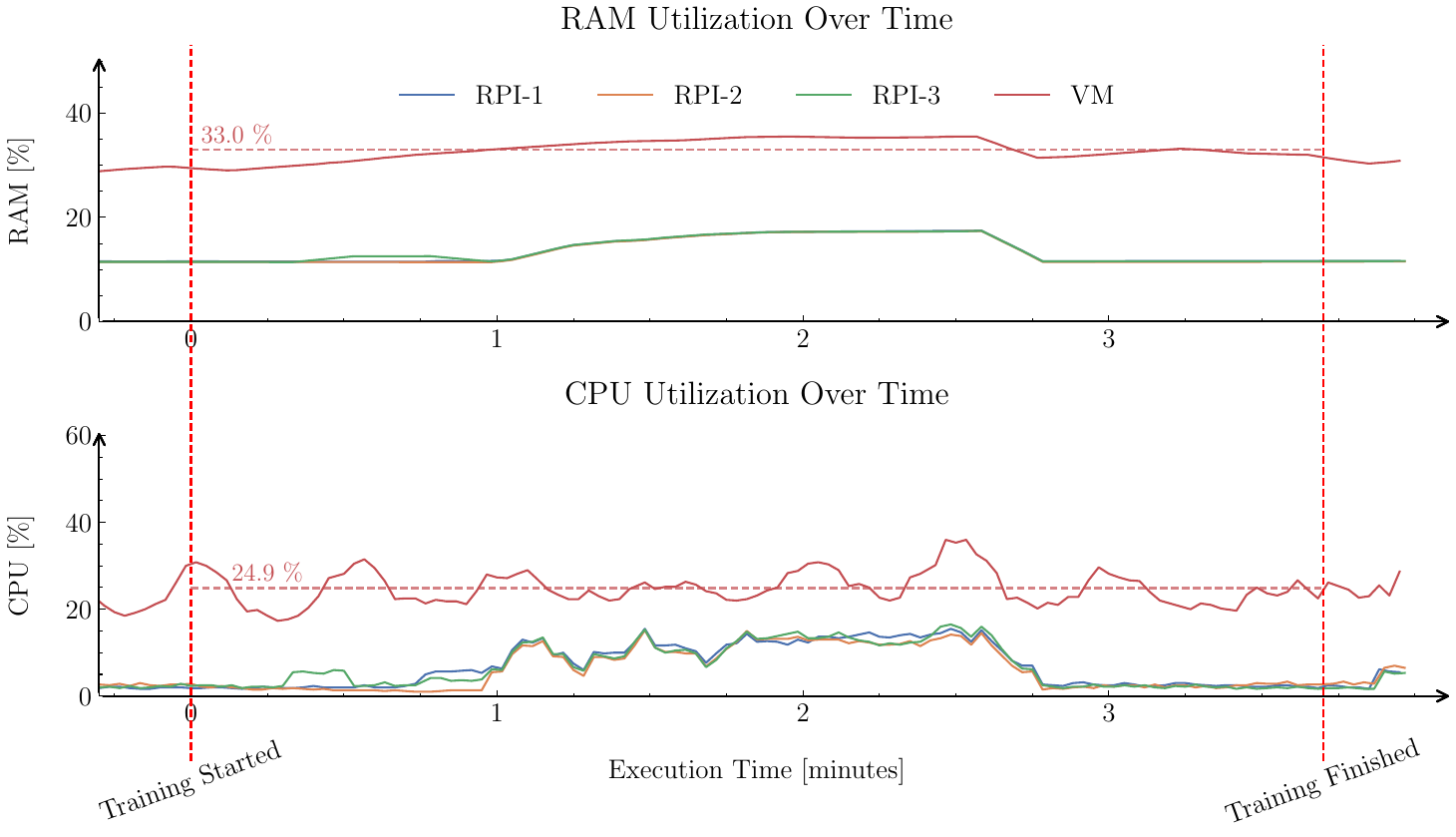}
        \caption{Distributed training (Scenario 3).}
        \label{subfig:distributed_training}
        \vspace{10pt}
    \end{subfigure}
    \caption{Resource utilization during model training.}
    \label{fig:resource_utilization_training}
\end{figure}

From Fig.~\ref{fig:resource_utilization_training}, we can see that for smaller batch sizes the distributed training performed considerably worse.
For batch size 8, the centralized training required 216 seconds while the distributed training required 772 seconds, therefore the centralized executed $\approx$$72\%$ faster compared to the same number of CPU cores when training on a distributed setup.
As the batch size increased, the two scenarios started converging.
At the largest tested batch size (512), a centralized training with Scenario 1 executed 15\% faster ($\Delta 15 s$) than distributed training with Scenario 3.
From Fig.~\ref{fig:resource_utilization_training}, we can also see that for batch sizes larger than 16, Scenario 3 on 4 nodes with 1 CPU core outperforms the Scenario 2 limited to one CPU.
At the largest batch size, distributed training takes 199 seconds, while training with Scenario 2 takes 354 seconds, therefore distributed training executes in $\approx$$44\%$ less time than training on 1 CPU core, proving the benefits of distributed training compared to the baseline.

The O-RAN model training performs similarly to the NAOMI with Scenario 1, however, as explained in Section~\ref{sec:exp-setup}, the O-RAN model training component can utilize a larger amount of CPU.

Fig.~\ref{fig:resource_utilization_training} depicts RAM and CPU utilization during QoE model training with NAOMI.
Figure~\ref{subfig:single_node_training} illustrates training with Scenario 1, while Figure~\ref{subfig:distributed_training} illustrates training Scenario 3, as explained in Section~\ref{sec:exp-setup} and summarized in Table~\ref{tab:scenarios_setups}.
Resource utilization for all nodes (edge nodes and VM) in a distributed setup is plotted in Fig.~\ref{subfig:distributed_training}. 

With Scenario 1, depicted in Fig.~\ref{subfig:single_node_training}, the average RAM utilization is $35.2\%$, while the CPU utilization is $30.1\%$.
When performing distributed training with Scenario 3, presented in Fig.~\ref{subfig:distributed_training}, the average RAM utilization is $33\%$, while the CPU utilization is $24.9\%$ on the virtual machine (VM).

The average RAM and CPU utilization on the VM during the model training are 2 and 5 percentage points lower, respectively, for the distributed setup.
This indicates an advantage of distributed training for lowering resource usage on the main node by dispersing parts of the computation on edge devices and achieving similar performance as shown in the previous results.
At the same time, the resource usage on the edge devices (Raspberry Pi 5) does not exceed $20\%$ of the total node resources, showing that those devices can perform their primary tasks with the majority of their resources while cooperating in the model training.

\subsection{QoE Inference Performance}

\begin{figure}[htbp]
\centerline{\includegraphics[width=1\linewidth]{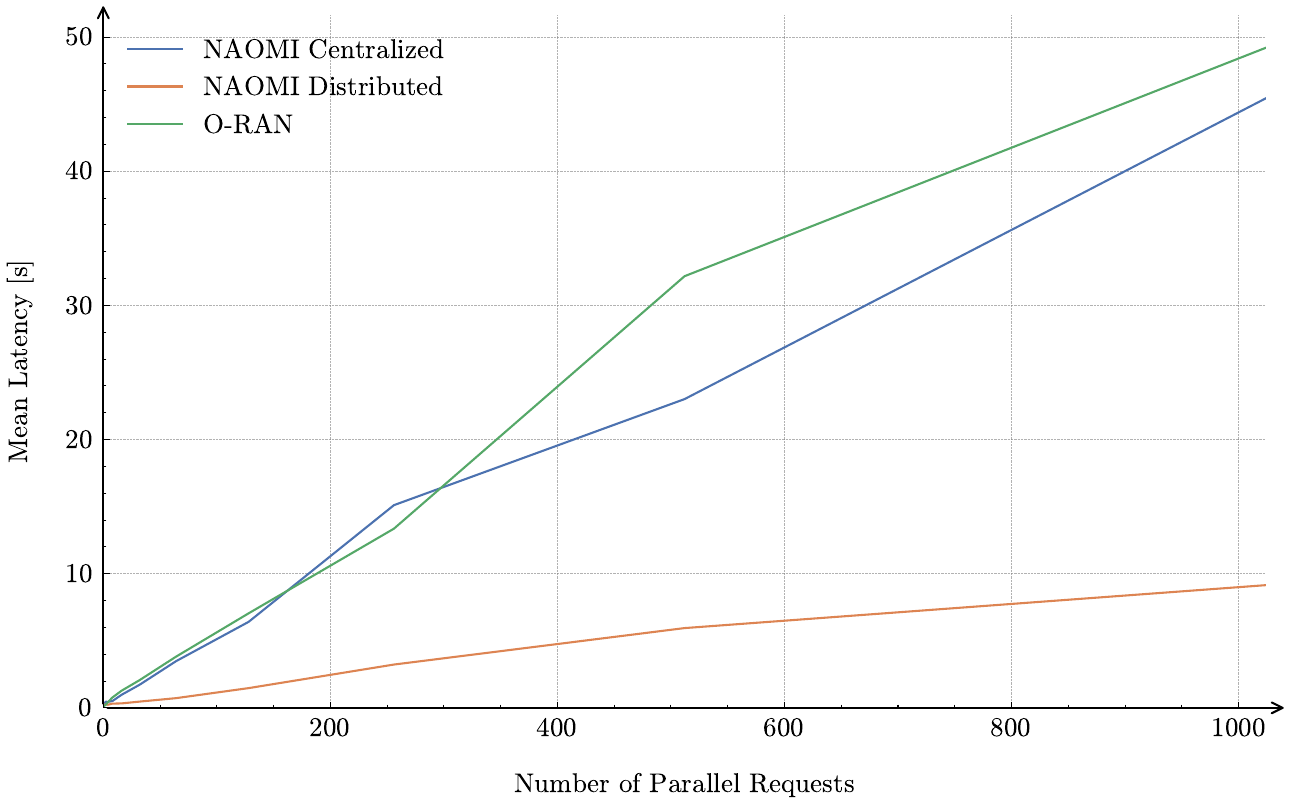}}
\caption{Mean end-to-end latency for concurrent inference API requests.}
\label{fig:lat}
\end{figure}

Fig.~\ref{fig:lat} presents mean end-to-end latencies for concurrent inference requests on the deployed QoE prediction model.
By increasing the number of concurrent requests (\(x\) axis) sent to the API endpoint, the mean latency increases (measured in seconds on \(y\) axis).
The mean latencies for the O-RAN solution as well as both NAOMI solutions, centralised and when deployed on a distributed infrastructure, are plotted on the graph.

From Fig.~\ref{fig:lat}, we can see that inference tools for both AI/ML workflows perform comparably up to 300 concurrent requests.
At 512 concurrent requests, the average end-to-end latency is 23 seconds for NAOMI on a centralized setup and 32 seconds for O-RAN.
For larger concurrent requests, the NAOMI inference service on a centralized setup has up to $28\%$ smaller mean end-to-end latency.
The configurations for both inference components were kept at default, so the differences are likely due to different horizontal autoscalers, which control the number of model replicas, as discussed in Section~\ref{subsec:methods:inference}.

On a distributed setup, there are 4 times more CPU cores available for model inference, as discussed in Section~\ref{subsec:methods:inference}.
From Fig.~\ref{fig:lat}, we can see that at 512 concurrent requests, the average end-to-end latency is 3.9 seconds on a distributed setup compared to 23 seconds on a centralized setup.
Calculating a speedup coefficient (coefficient of mean latencies between setups) shows a speedup between 1.2 and 3 for up to 32 concurrent requests and a speedup of 5.9 for 512 concurrent requests.
A lower speedup for a lower number of concurrent requests is due to a smaller number of replicas, as horizontal autoscaling creates new replicas when the load increases.
Speedup above 4 is possible because of load-balancing on multiple nodes and network congestion when running inference on a single node, as well as having a different CPU architecture on edge nodes.
These results show the ability of NAOMI to scale the inference service on a distributed edge network.

\subsection{Heterogeneous Scale Study}
\label{subsec:results:scale_study}

\begin{figure}[htbp]
\centerline{\includegraphics[width=1\linewidth]{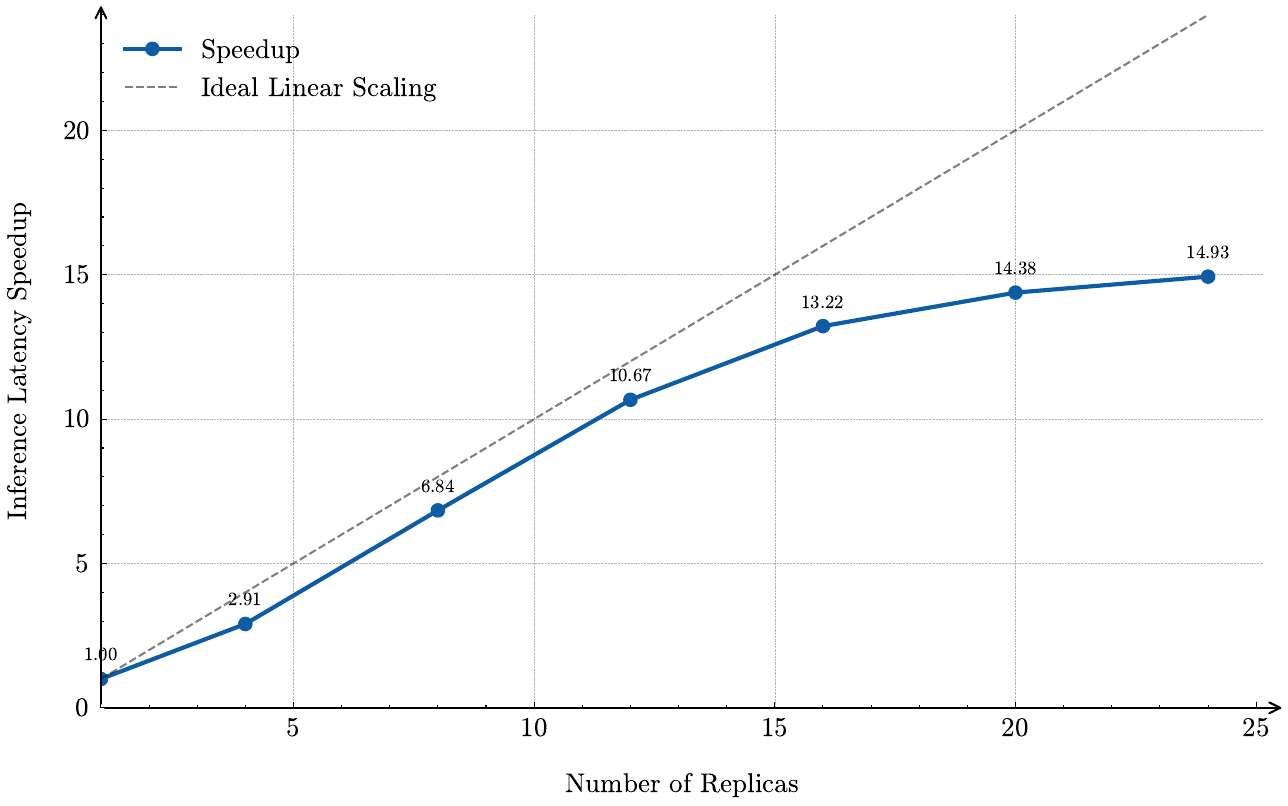}}
\caption{Inference scale study on distributed heterogeneous nodes.}
\label{fig:scale_study}
\end{figure}

Fig.~\ref{fig:scale_study} depicts mean inference latency speedup compared to the increasing number of model replicas when running 500 concurrent requests on a deployed QoE model. As discussed in Section~\ref{subsec:methods:scale_study} by increasing the number of replicas a linear latency reduction is expected.
The ideal linear speedup is plotted on the graph for comparison.

As we can see from Fig.~\ref{fig:scale_study}, the speedup increases when increasing the number of replicas. The speedup at 4 replicas is 2.9, and at 8 replicas, it is 8.8. The variations are expected due to heterogeneous architecture, meaning some CPU architectures will perform faster inference on the same amount of CPU cores, so the results will depend on where the models are placed.
We can also see that up to 12 replicas, the speedup increases nearly linearly, and at 16, 20, and 24 model replicas, we can observe diminishing returns, as the speedup at 20 replicas is 14.4 and at 24 replicas is 14.9, showing only a slight advantage of having more replicas.

The results show the ability for inference service to scale on NAOMI, however for the chosen model there are no significant benefits of increasing the number of model replicas above 16. Benefits at a larger number of model replicas are expected for models with longer inference latency or for larger amounts of concurrent requests.

\subsection{Workflows Beyond O-RAN}
\label{subsec:resoults:general_models}

\begin{figure}[htbp]
\centerline{\includegraphics[width=1\linewidth]{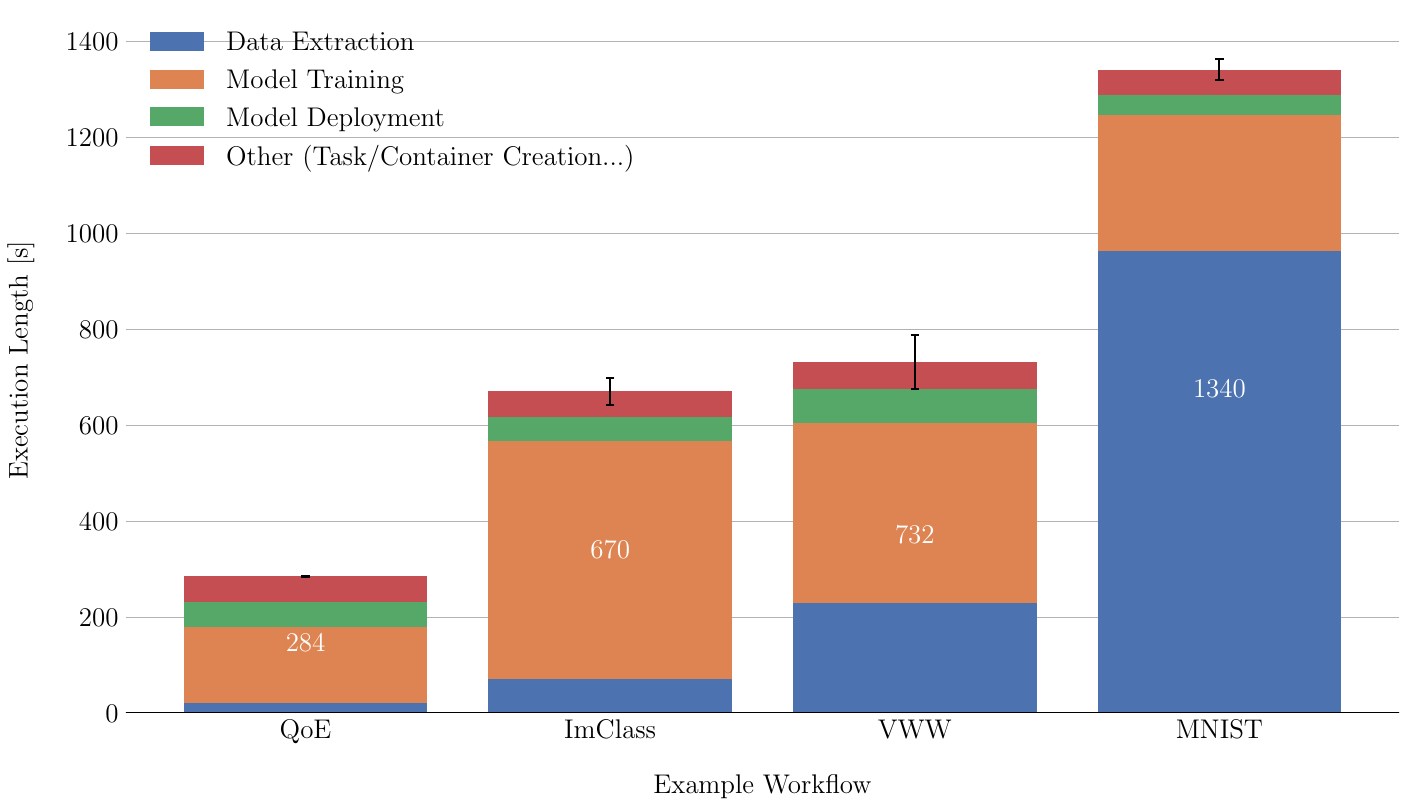}}
\caption{Execution time of example ML workflows.}
\label{fig:tinyML_exec_time}
\end{figure}

Fig.~\ref{fig:tinyML_exec_time} presents execution times of four workflows examples: QoE (Quality of Experience) from O-RAN, ImClass (Image Classification), VWW (Visual Wake Words) from MLPerf TinyML benchmark, and MNIST handwritten digit classification as presented in Section~\ref{subsec:methods:general_models}. Each column in Fig.~\ref{fig:tinyML_exec_time} consists of sections that present the execution time of different workflow stages. Experiments were repeated 10 times and the results present the average of those experiments.

QoE workflow example with $284 \pm 1$ second takes, on average, the least time to complete. ImClass and VWW take $670 \pm 28s$ and $732 \pm 57s$ respectively, while the MNIST workflow example takes $1340 \pm 21$ seconds.

Data extraction requires 21 seconds for the QoE example, while ImClass takes, on average, 71 seconds. On average, VWW requires 229 seconds and MNIST 962 seconds. 
Considering datasets are of different sizes and have different numbers of samples as presented in Section~\ref{subsec:methods:general_models}, variation in the results is expected. As we can see in Fig.~\ref{fig:tinyML_exec_time}, model training and data extraction are not correlated.
Training time to data extraction time ratios are approximately 7.5 for QoE (Meaning model training takes 7.5 times longer than data extraction), 7.0 for ImClass, 1.6 for VWW, and 0.3 for MNIST. 
Differences in ratios are due to model training depending on model complexity, the different dataset sizes and different data extraction methods, as discussed in Section~\ref{subsec:methods:general_models}. 
For model training, QoE took 158 seconds on average, MNIST required 285s, VWW 373s, and the ImClass required the most time for model training, finishing after 495 seconds on average.

Model deployment varied between example workflows, with MNIST finishing at 41 seconds on average and VWW taking the most time around 73 seconds on average.
\textit{Other} (overhead) tasks did not differ significantly between different workflow examples at $\approx$$53$ seconds on average. 
Overall, the results show that data extraction and model training contribute the most to the total execution time, while deployment and overhead tasks remain relatively stable across all workflows.

\begin{figure}[htbp]
\centerline{\includegraphics[width=1\linewidth]{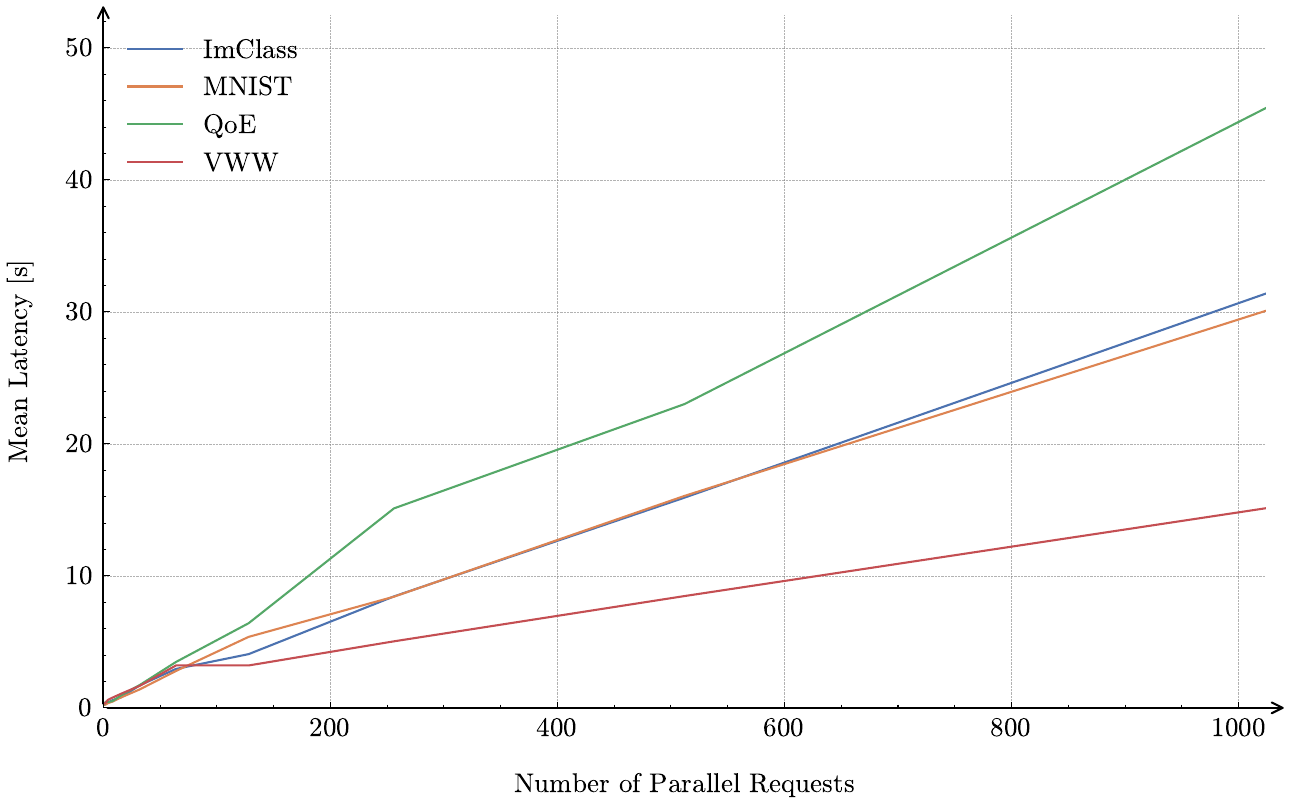}}
\caption{Mean end-to-end latency for concurrent inference API requests.}
\label{fig:tinyML_inference}
\end{figure}

Fig.~\ref{fig:tinyML_inference} depicts mean end-to-end latencies for concurrent inference requests on the deployed example workflow models and it reveals that that different models perform comparably up to 50 parallel requests on average, with the difference at 32 parallel requests less than 0.3s between each two models.

At 200 parallel requests and higher, ImClass and MNIST models perform similarly, with the difference at above 1000 parallel requests 1.3s on average. QoE model required $\approx$$45$ seconds to process 1024 parallel requests on average and VWW model required $\approx$$15$ seconds on average at the same number of parallel requests.

Concurrent requests affect response times differently across each model, yet they remain similar at moderate loads, with the differences becoming more noticeable at higher loads.  
MNIST and VWW models performed similarly.  
Among MLPerf TinyML models, VWW outperforms ImClass, which is consistent with the results from the paper discussing the proposed benchmarks and models~\cite{banbury2021mlperf}. 
Model inference latency is heavily dependent on the model complexity. 
However, these parallel requests are sent over the network, so some communication overhead contributing to the differences is expected.

\subsection{Discussion of Democratization Requirements}
\label{subsec:eval_requirements}

\begin{table*}[htbp]
\caption{Summary of Supported Requirements by AI/ML workflow solutions}
\ra{1.2}
\centering
\fontsize{8pt}{10pt}\selectfont 
\begin{tabularx}{1\linewidth}{@{}>{\centering\arraybackslash}p{3cm}|>{\centering\arraybackslash}X|>{\centering\arraybackslash}X@{}}
\toprule
\thead{\textit{\textbf{Requirements}}} & \textit{\textbf{NAOMI}} & \textit{\textbf{O-RAN}} \\
\midrule\midrule
\textbf{Openness} & Source code is available on Github\textsuperscript{1} & Source code is available on O-RAN Gerrit\textsuperscript{2} \\
 \midrule
\textbf{Virtualization} & Containerized components, Kubernetes pods & Containerized components, Kubernetes pods \\
\midrule
\textbf{Distributed Services} & Multi node Kubernetes & Multi node Kubernetes \\
\midrule
\textbf{Scalability} & Kubernetes autoscaling & Kubernetes autoscaling \\
\midrule
\textbf{Ease of Use} & 3 steps to deployment & More than 10 steps to deployment \\
& 3 steps to run the example workflow & More than 15 steps to run the example workflow  \\
& 1 configuration file & Configurations are in multiple places \\
& Requirements file & Requirements are part of the installation procedure \\
& 4 example workflows provided & 1 example workflows provided \\
\midrule
\textbf{Modularity} & 5 components & 5 components  \\
& 4 AI/ML components can be disabled or enabled and utilized separately  &  Deployed as one entity with 2 optional components \\
& Modules can be replaced and are not interdependent & Most modules can not be replaced \\
\midrule
\textbf{Self-Evolving} & Monitoring is supported & No monitoring metrics are collected or displayed \\
& Retraining and model deployment are automated & Retraining and model deployment are triggered manually \\
& Workflows can be triggered based on monitoring metrics &  \\
\midrule
\textbf{Heterogeneity} & Supports ARM and x86 architectures & Supports x86 architecture \\
& Deployment on 1 x86 VM and 3 ARM Raspberry Pi devices & Deployment on 1 x86 device \\
\bottomrule
\multicolumn{3}{l}{\textsuperscript{1}\url{https://github.com/copandrej/NAOMI} \quad
\textsuperscript{2}\url{https://gerrit.o-ran-sc.org/r/admin/repos/}}
\end{tabularx}
\label{tab:req_summary}
\end{table*}

As outlined in Section~\ref{subsec:support_for_dem} and summarized in Table~\ref{tab:req}, NAOMI fulfills all eight democratization requirements for AI/ML workflow systems, identified in Section~\ref{sec:requirements}, while the O-RAN implementation supports four. 
We argue that NAOMI supports the democratization requirements of ease of use, modularity, self-evolving, and heterogeneity to a greater extent than O-RAN; moreover, in some cases, the O-RAN solution does not support the democratization requirements at all.
We evaluate these claims through quantitative metrics such as the number of steps and configurations and architectural analysis focused on provided documentation~\cite{o-ran-sc-docs} and hands-on experience, with main findings summarized in Table~\ref{tab:req_summary} and further discussed in the following sections.

\subsubsection{Ease of use}

As we can see in row \textit{Ease of Use} in Table~\ref{tab:req_summary}, NAOMI requires developers to follow three steps to deploy the solution on their premises. 
These are installing the Kubernetes cluster using an installation script, deploying NAOMI with Helm, and configuring the local environment with a script. 
On the other hand, O-RAN requires more than 10 steps with multiple commands in each step~\cite{O-RANSoftwareCommunity2024O-RANDocumentation}.

Similarly, three steps are required for running the example workflow: populating the database with data, starting the Flyte workflow, and monitoring the progress. 
O-RAN requires more than 15 steps to achieve the same progress on their example workflow.

Further, all NAOMI configurations are in one configuration file, which configures helm charts for all modules and the requirements file for local packages is presented in the repository. On the other hand, O-RAN does not have a central configuration or requirements file. 
The deployment instructions specify what to configure and which packages to install. To simplify the process of adapting workflows to our system, NAOMI includes 4 example workflows, while O-RAN includes one.

\subsubsection{Modularity}

The sixth row in Table~\ref{tab:req_summary} presents the support of the modularity democratization requirement for both solutions. It can be seen that both solutions consist of 5 components with possibly multiple tools for each component as discussed in Section~\ref{sec:methodology} and listed in Table~\ref{tab:components}.

While both solutions consist of components with similar functionality, modularity is supported by NAOMI as components can be enabled and disabled separately, based on the user needs thus also enabling saving on resource utilization when needed.
For example, with only \textit{MLFlow} model store enabled, NAOMI requires 4 CPU cores and 8 GB of RAM to operate. 
Further, modules can be replaced as they are not interdependent, whereas all components are connected with interfaces in O-RAN.

\subsubsection{Self-Evolving}

As presented in Section~\ref{sec:requirements}, monitoring and automated retraining are crucial for self-evolving solutions. In Table~\ref{tab:req_summary} in row \textit{Self-Evolving}, we can see that NAOMI supports monitoring through its monitoring tools Prometheus and Grafana and provides a solution for scheduled retraining based on monitoring metrics using Flyte workflow orchestrator. 
The example self-evolving workflow is provided in the NAOMI repository with jobs running the pipeline every 30 minutes if Prometheus metrics show that the deployed model does not have a status of healthy.

On the other hand, O-RAN retraining and model deployment are manual, as users are required to start the training job and deploy the model using multiple commands.

\subsubsection{Heterogeneity}

Table~\ref{tab:req_summary} in row \textit{Heterogeneity} summarizes the support for heterogeneity requirements by NAOMI and O-RAN solution. 
NAOMI supports deployment on ARM and x86 architectures. Specifically, it can be deployed on multiple Raspberry Pi 5 nodes and a virtual machine with an Intel CPU, as discussed in Section~\ref{sec:exp-setup}. 
O-RAN solution supports x86 architecture and was successfully deployed on one virtual machine. 
These two deployments were evaluated on an example workflow in Section~\ref{sec:results} showing that NAOMI supports the heterogeneity democratization requirement.

\section{Conclusion} \label{sec:conclusion}

In this paper, we propose \textit{NAOMI}, a solution for democratizing AI/ML workflows at the network edge. NAOMI is a democratized solution following the identified requirements of a democratized network edge, such as ease of use, deployment heterogeneity, and modularity and this way brings AI/ML workflow automation closer to the public and lowers the boundary for anyone to deploy and utilize it.

By analyzing the architecture of AI/ML workflow as defined by the O-RAN Alliance and comparing available open-source software, we have selected the SotA tools for the solution and proposed an architecture that can realize all essential AI/ML workflow components such as data preparation, model training, model management, model deployment, continuous operation, and monitoring. NAOMI can be deployed on a distributed heterogeneous cluster comprising general-purpose x86 and ARM edge devices, providing a modular, scalable, distributed, and architecture-independent solution for the network edge.

By evaluating and comparing it to the latest release of the AI/ML Framework built by the O-RAN Software Community, we showed that NAOMI performs on par with model inference and training, while NAOMI is deployed 40\% faster and the example Quality of Experience workflow executes up to 73\% faster for larger datasets.
We highlighted the advantages of distributed training in spreading resource usage across multiple nodes while achieving similar performance to single-node training on an equivalent number of CPU cores.
Furthermore, distributing ML models across multiple edge and cloud nodes results in model inference improvements consistent with theoretical speedup.

In our future work, we aim to enhance NAOMI's reliability and flexibility, leveraging its distributed capabilities to explore advanced model training techniques. We will focus on utilizing the power of GPU-enabled edge devices for accelerated model training and inference, and develop intelligent orchestration mechanisms to optimally distribute computational tasks like model inference across the network edge.

\section*{Acknowledgment}
This work was supported by the Slovenian Research Agency (P2-0016) and the European Commission NANCY project (No. 101096456).

\bibliographystyle{elsarticle-num} 
\bibliography{references, mendeley-ref}


%



\end{document}